\documentclass[11pt]{article}

\usepackage{amsmath,amsfonts,amssymb,graphicx,psfrag,rotating}

\usepackage[top=1in, left=1in, right=1in,bottom=1in]{geometry}   % See geometry.pdf to learn the layout options. There are lots.

\usepackage{setspace} 
\usepackage{dcolumn} 
\usepackage{natbib}
\bibpunct{(}{)}{;}{a}{}{,}
\usepackage{comment}
\usepackage{rotate}
\usepackage{multirow}
\usepackage{lscape}
\usepackage{bbm}
\usepackage{scalefnt}
\usepackage[autolanguage]{numprint}
\usepackage{paralist}
\usepackage{caption}
\usepackage{hyperref}
\hypersetup{
	colorlinks   = true,
	linkcolor = red,
	citecolor    = blue
}
\usepackage{booktabs}
\usepackage{bm}
\usepackage[compact]{titlesec}

\usepackage{natbib}
\bibpunct{(}{)}{;}{a}{}{,}

\usepackage{xr}
\makeatletter

\newcommand*{\addFileDependency}[1]{% argument=file name and extension
\typeout{(#1)}% latexmk will find this if $recorder=0
\@addtofilelist{#1}
\IfFileExists{#1}{}{\typeout{No file #1.}}
}\makeatother

\usepackage{tikz}
\usetikzlibrary{trees}
\usepackage{appendix}
\usepackage{threeparttable}

\usepackage{caption}
\usepackage{subcaption}

\usepackage{lmodern}
\fontfamily{lmtt}\selectfont
\usepackage[T1]{fontenc}

\titleformat{\section}
  {\normalfont\fontsize{11}{15}\bfseries}{\thesection}{1em}{}

\titleformat{\subsection}
  {\normalfont\fontsize{11}{15}\bfseries}{\thesubsection}{1em}{}

\usepackage{natbib}
\usepackage{amsmath,amsthm, amssymb, bbm}
\usepackage{graphicx}
\usepackage{mathtools}
\usepackage{color}
\usepackage{dcolumn}
\newcolumntype{.}{D{.}{.}{-1}}
\newcolumntype{d}[1]{D{.}{.}{#1}}

\newtheorem{assumption}{Assumption}

\newcommand{\bbb}{\boldsymbol \beta}

\newcommand{\one}{\mathbf 1}

\newcommand{\bo}{\boldsymbol o}

\newcommand{\bg}{\boldsymbol g}

\newcommand{\br}{\boldsymbol r}

\newcommand{\be}{\boldsymbol e}

\newcommand{\bX}{\mathbf X}

\newcommand{\bZ}{\mathbf Z}
\newcommand{\bD}{\mathbf D}
\newcommand{\bz}{\mathbf z}

\newcommand{\bU}{\mathbf U}

\newcommand{\bV}{\mathbf V}

\newcommand{\bW}{\mathbf W}
\newcommand{\bw}{\mathbf w}

\newcommand{\bb}{\mathbf b}

\newcommand{\bu}{\boldsymbol u}
\newcommand{\bv}{\boldsymbol v}
\newcommand{\bx}{\boldsymbol x}

\newcommand{\ME}{\mathbb E}
\newcommand{\MP}{\mathbb P}

\newcommand\independent{\protect\mathpalette{\protect\independenT}{\perp}}
\def\independenT#1#2{\mathrel{\rlap{$#1#2$}\mkern2mu{#1#2}}}

\usepackage{amsthm}
\theoremstyle{plain}

\newtheorem{theorem}{Theorem}

\newtheorem{algorithm}{Algorithm}
\usepackage{tikz}
\usetikzlibrary{
	graphs,
	graphs.standard
}

\def\spacingset#1{\renewcommand{\baselinestretch}%
{#1}\small\normalsize} \spacingset{1}

\begin{document}

\pagestyle{plain}
%%%%%%%%%%%%%%%%%%%%%%%%%%%%%%%%%%%%%%%%%%%
\newcommand{\blind}{0}

\newcommand{\tit}{\bf Flexibly Estimating and Interpreting Heterogeneous Treatment Effects of Laparoscopic Surgery for Cholecystitis Patients }

\if0\blind

{\title{\tit\thanks{The authors declare no conflicts. Research in this article was supported by the National Library of Medicine, \#1R01LM013361-01A1. All statements in this report, including its findings and conclusions, are solely those of the authors. The dataset used for this study was purchased with a grant from the Society of American Gastrointestinal and Endoscopic Surgeons. Although the AMA Physician Masterfile data is the source of the raw physician data, the tables and tabulations were prepared by the authors and do not reflect the work of the AMA. The Pennsylvania Health Cost Containment Council (PHC4) is an independent state agency responsible for addressing the problems of escalating health costs, ensuring the quality of health care, and increasing access to health care for all citizens. While PHC4 has provided data for this study, PHC4 specifically disclaims responsibility for any analyses, interpretations or conclusions. Some of the data used to produce this publication was purchased from or provided by the New York State Department of Health (NYSDOH) Statewide Planning and Research Cooperative System (SPARCS). However, the conclusions derived, and views expressed herein are those of the author(s) and do not reflect the conclusions or views of NYSDOH. NYSDOH, its employees, officers, and agents make no representation, warranty or guarantee as to the accuracy, completeness, currency, or suitability of the information provided here. This publication was derived, in part, from a limited data set supplied by the Florida Agency for Health Care Administration (AHCA) which specifically disclaims responsibility for any analysis, interpretations, or conclusions that may be created as a result of the limited data set.}}
\author{Matteo Bonvini\thanks{Assistant Professor, Rutgers University, Email: mb1662@stat.rutgers.edu}
\and Zhenghao Zeng\thanks{Carnegie Mellon University, Email: zhenghaz@andrew.cmu.edu}
\and Miaoqing Yu\thanks{University of California, Santa Barbara. Email: miaoqingyu@umail.ucsb.edu}
\and Edward H. Kennedy\thanks{Associate Professor, Carnegie Mellon University, Email: edward@stat.cmu.edu}
\and Luke Keele\thanks{Associate Professor, University of Pennsylvania, Email: luke.keele@gmail.com, corresponding author}
}

\date{\today}

\maketitle
}\fi

\if1\blind
\title{\bf \tit}
\maketitle
\fi

\begin{abstract}
Laparoscopic surgery has been shown through a number of randomized trials to be an effective form of treatment for cholecystitis. Given this evidence, one natural question for clinical practice is: does the effectiveness of laparoscopic surgery vary among patients? It might be the case that, while the overall effect is positive, some patients treated with laparoscopic surgery may respond positively to the intervention while others do not or may be harmed. In our study, we focus on conditional average treatment effects to understand whether treatment effects vary systematically with patient characteristics. Recent methodological work has developed a meta-learner framework for flexible estimation of conditional causal effects. In this framework, nonparametric estimation methods can be used to avoid bias from model misspecification while preserving statistical efficiency. In addition, researchers can flexibly and effectively explore whether treatment effects vary with a large number of possible effect modifiers. However, these methods have certain limitations. For example,  conducting inference can be challenging if black-box models are used. Further, interpreting and visualizing the effect estimates can be difficult when there are multi-valued effect modifiers. In this paper, we develop new methods that allow for interpretable results and inference from the meta-learner framework for heterogeneous treatment effects estimation. We also demonstrate methods that allow for an exploratory analysis to identify possible effect modifiers. We apply our methods to a large database for the use of laparoscopic surgery in treating cholecystitis. We also conduct a series of simulation studies to understand the relative performance of the methods we develop. Our study provides key guidelines for the interpretation of conditional causal effects from the meta-learner framework. 

\end{abstract}

\noindent%

\thispagestyle{empty}

\clearpage

\spacingset{1.59}

\section{Introduction}

Comparative effectiveness research (CER) seeks to identify interventions that improve healthcare outcomes. In our study, we focus on the comparative effectiveness of laparoscopic surgery compared to open surgery for patients with cholecystitis. In CER studies, investigators typically focus on the average treatment effect (ATE), i.e., how an intervention changes patient outcomes on average, at the population level. For example: how would the average patient fare if they opted for laparoscopic surgery versus open surgery? While the ATE is a useful summary of a causal effect, it can mask important patient-to-patient variation \citep{ding2019decompose}. For example, the ATE could obscure the fact that some patients respond dramatically to treatment while others suffer adverse reactions. 

Conditional average treatment effects (CATEs), in contrast, describe how the effects vary with measured features. CATEs can be crucial to study for a number of reasons, including: (i) to design optimal treatment policies, (ii) to improve our understanding of systematic variation in treatment effects, and (iii) to help inform how future treatments should be developed (i.e., how to target those who do not benefit from currently available treatments). In summary, while the ATE is a useful starting point for understanding treatment effectiveness, CATEs are crucial component for understanding how effects vary among patients and for tailoring treatment strategies.

\subsection{Laparoscopic surgery for cholecystitis}

We perform an original study on the effectiveness of laparoscopic surgery (LS) for the treatment of cholecystitis.  Cholecystitis is a disease that results from inflammation of the gallbladder. One common cause of cholecystitis is gallstones, which block the tube leading out of gallbladder. Cholecystitis is also caused by bile duct problems, tumors, serious illness and other types of infections. It can lead to infections or inflammation of the pancreas (pancreatitis). Serious cholecystitis is often treated as an emergency condition when pain becomes acute and patients seek treatment in the emergency room. Treatment for cholecystitis is often done via either two kinds of surgery: laparoscopic or open. Laparoscopic surgery (LS) is a surgical technique designed to create smaller incisions than standard open surgery (OS). More precisely, under laparoscopic surgery small narrow tubes--called trochars--are inserted into the abdomen through small (less than one centimeter) incisions. Through these trochars, clamps, scissors, and sutures are inserted, and the surgeon uses these instruments to manipulate, cut, and sew tissue. A camera, inserted through one trochar, is linked to a video monitor, which allows the surgeon to view the abdominal contents. The use of smaller incisions is designed to help patients recover more quickly and experience fewer surgical complications. 

A large number of randomized controlled trials (RCTs) have been conducted in this area.  See \citet{johansson2005randomized,loozen2018laparoscopic,madureira2013randomized} as well-known examples. One meta-analysis of ten randomized trials concluded that LS resulted in lower complications and death rates compared to OS \citep{coccolini2015open}. Thus, the consensus in the clinical literature is that the ATE for LS is known and positive in the sense that, on average, patients will experience better outcomes if they receive LS for cholecystitis.  However, it is also likely that the effect of LS varies from patient to patient. More specifically, there may be some patients for whom LS is particularly beneficial, and there may be other patients for whom it is harmful. The magnitude of this variation is not fully understood and it is the central goal of our work.

We analyze an observational dataset aiming to evaluate the relative effectiveness of open surgery for cholecystitis compared to laparoscopic surgery with a focus on estimating CATEs. We seek to replicate the results from the RCTs using a large observational data set that allows us to better detect whether the effects of LS vary systematically with key patient characteristics. Specifically, we use a data set that merges the American Medical Association (AMA) Physician Masterfile with all-payer hospital discharge claims from New York, Florida, and Pennsylvania in 2012-2013. In this dataset, there are 116,234 patients that underwent a cholecystectomy, of whom 94,485 underwent LS and 21,749 underwent OS. Our primary outcome is an adverse outcome after surgery, which is defined as the presence of either a post-operative complication or a prolonged length of stay (PLOS). PLOS is measured using an indicator variable which equals one when the hospital and operation-specific length of stay is greater than the 75th percentile. PLOS serves as a generalized measure for a variety of post-operative complications that may not be specifically captured in claims data \citep{bansal2016using}. The data also includes an identifier for patient, patient sociodemographic, clinical characteristics including indicators for frailty, severe sepsis or septic shock, and 31 comorbidities based on Elixhauser indices \citep{elixhauser1998comorbidity}, as well as information on insurance type. In addition, we generate a risk score for each patient \citep{silber2016comparison} and use it as an additional baseline covariate. The risk score is a prediction from a logistic model estimated in the following way. First, we randomly sample 2\% of the study population and, within this sub-sample, we regress the adverse event outcome on all baseline covariates using a logistic regression model with main effects. Risk scores for the remaining 98\% patients are generated as predictions of their probability of an adverse event from this model. The patients in the sub-sample are discarded and not used for further analysis. 

We use the baseline covariates not only as confounders to adjust for differences in the treated and control populations but also as possible effect modifiers, i.e., variables that modify the effect of LS. In our study, as is typical in many clinical investigations, there is a small subset of these variables that clinical expertise identifies as effect modifiers \emph{a priori}. That is, covariates such as sepsis, age, and predicted risk are, based on clinical expertise, thought to be key effect modifiers. For example, LS may be more effective for septic or older patients that are less able to withstand the invasive nature of OS. However, with over 40 baseline covariates, there may be other effect modifiers that we would like to identify \emph{a posteriori}.

\subsection{Methods for Estimating Heterogeneous Causal Effects}

The most common approach for estimating CATEs relies on simple parametric models, e.g., linear or generalized linear models with multiplicative terms. However, the use of parametric models for estimating CATEs is typically done based on convenience or convention, rather than actual substantive knowledge of underlying mechanisms. When parametric models are misspecified the resulting estimates will generally be biased, potentially severely. For example, we may expect the effect of treatment to vary with risk in a nonlinear way. Standard parametric models are not readily able to provide accurate estimates of CATE with complex functional forms.  

Of late, there has been considerable interest in adapting flexible machine learning (ML) methods to the problem of estimating CATEs. In particular, the meta-learner framework has been developed to offer a principled way to deploy ML methods for estimating conditional treatment effects \citep{kunzel2019metalearners, foster2023orthogonal, kennedy2020towards,nie2021quasi, morzywolek2023general}. The expression ``meta-learner'' refers to the fact that these methods describes algorithms or procedures to estimate treatment effects that are agnostic with respect to the specific ML method(s) the analyst will choose. A central goal of this framework is to construct estimators enjoying the desirable properties of influence-functions-based estimators of pathwise differentiable parameters, such as double-robustness and second-order bias. The outcomes of these learners are typically estimates of \emph{individual} treatment effects (ITEs), which are conditional average treatment effects defined in terms of the whole vector, of potentially high dimension, of baseline covariates. For these methods to provide interpretable results in applied studies, additional methodological development is needed.

\subsection{Our Contribution: Methods for Interpretation of Estimated Conditional Causal Effects}

In this work, we propose methods for visualizing and interpreting treatment effect heterogeneity in the meta-learner framework. We focus on the DR-learner framework for estimating conditional average treatment effects (CATEs) \citep{kennedy2020towards}. We outline several aspects that one should consider when estimating CATEs. In the presence of several continuous effect modifiers, the full CATE curve may be hard to interpret. We therefore consider three ways to improve visualization and conduct straightforward inference. First, we discuss pros and cons of considering univariate CATEs, i.e., treatment effect curves defined with respect to a single effect modifier of interest. This measure of effect heterogeneity, while intuitive and well-defined, may nonetheless capture heterogeneity that is not directly attributable to the effect modifier of interest. This can happen, for example, whenever such modifier is correlated with other covariates. To circumvent this potential issue, we consider imposing an additive assumption on the full CATE curve, aiming to incorporate all effect modifiers in the definition of the CATE curve while preserving interpretability. Because such additivitiy assumption might be hard to justify in practice, we also consider estimating a partial dependence function, which is a univariate function of the effect modifier of interest that nevertheless accounts for the heterogeneity attributable to the other effect modifiers as well. In all cases, we outline methods for constructing confidence bands. Finally, we demonstrate how one can perform an exploratory analysis to identify possible effect modifiers not known \textit{a priori}. We use a variable importance metric studied in \cite{hines2022variable} to identify a set of possible effect modifiers among a candidate set of variables. In short, we outline an analytic framework for testing an \emph{a priori} specification of effect modifiers and an \emph{a posteriori} analysis to discover additional potential effect modifiers. We conduct a series of simulation studies to understand the properties of the methods we have considered. Throughout, we integrate our methodological developments with our study of LS treatment for cholecystitis. We find that age, risk and sepsis are important effect modifiers. In addition, we find the CATEs for age and risk vary in a nonlinear fashion, which highlights the need to use flexible estimation methods in this context. Moreover, our exploratory analysis suggests that several other variables are also important effect modifiers. 

\section{Estimation of Average Treatment Effects} 
\label{sec:ATE}

In this section, we review state-of-the-art methods to efficiently estimate the average effect of LS vs OS for patients with cholecystitis by leveraging on modern machine learning methods. This section also introduces key statistical concepts needed to define estimators of treatment effects heterogeneity discussed in later sections.

\subsection{Notation, Estimands, and Assumptions}\label{sec:notations}

We use $A_i$ to denote a binary treatment for $i$-th individual. In our study, $A_i = 1$ indicates that the $i$-th individual receives LS and $A_i = 0$ indicates that they receive OS. The outcome is denoted as $Y_i$, with  $Y_i = 1$ indicating that the patient experiences an adverse outcome after surgery.
Let $\bX$ denote a set of baseline covariates that describe the units in the study prior to the treatment. In our study, $\bX$ includes measures of age, the number of comorbidities, our estimated risk score, an indicator for a disability, and indicators for a large number of possible comorbidities. We define $Y^a$ to be the potential outcome observed if the individual receives treatment $A = a$ \citep{rubin1974estimating, imbens2015causal}.  The average treatment effect (ATE) in the population is defined as $ \ME(Y^1 - Y^0)$, i.e., the expected difference in outcomes if every subject in the study population receives LS versus OS. The ATE, like all causal estimands, requires certain assumptions for identification. A set of assumptions commonly invoked is the following. 
\begin{assumption}\label{assumption:main}
    We assume that
    \begin{enumerate}
        \item $Y = Y^a \text{\, if \,} A=a$;
        \item There is sufficient \textit{overlap} in the covariates distribution between LS and OS groups, i.e., $0 < \MP(A=1 \mid \bX) < 1$;
        \item  The treatment assignment is \textit{ignorable} given measured covariates $\bX$, i.e., $(Y^1, Y^0) \independent A \mid \bX$.
    \end{enumerate}
\end{assumption}
The first condition is often referred to as the consistency or Stable Unit Treatment Value Assumption (SUTVA) \citep{Rubin:1986} in the literature. The third assumption requires that, for patients with similar covariate profiles, assignment to LS versus OS is as good as randomized. In our setting, since treatment is not randomly assigned, this assumption is not testable. However, we will evaluate its plausibility by comparing our estimates to those from randomized trials. 

Under Assumption \ref{assumption:main}, the CATE in terms of the entire covariate set $\bX$ is nonparametrically identified by:
\begin{align}
\ME(Y^1 - Y^0 \mid \bX) = \ME(Y \mid A=1,\bX) - \ME(Y \mid A=0, \bX).
\end{align}
and hence ATE is 
\begin{align}\label{eq:ate}
\ME(Y^1 - Y^0 ) = \ME \left\{\ME(Y \mid A=1,\bX) - \ME(Y \mid A=0, \bX) \right\}.
\end{align}

\subsection{DRML Estimation of the Average Treatment Effect}\label{sec:dr-ate}
The first step in our overall analysis of the effects of LS vs OS for patients with cholecystitis is to estimate the ATE as identified in Eq. \eqref{eq:ate}. To this end, we first provide a brief review of modern methods to efficiently estimate this parameter with machine learning. We refer to \cite{kennedy2022semiparametric} for a recent review on this subject. Let $\psi_a = \ME\{\ME(Y \mid A=a, \bX)\}$. In this section, we view $\psi_a$ as a functional of the unknown distribution of the data $\MP$. Its importance, however, rests on the fact that, under Assumption \ref{assumption:main}, this functional equals $\ME(Y^a)$, which is the mean potential outcome under treatment $A = a$. Here, we focus on estimating $\psi_1$ with the understanding that one can similarly estimate $\psi_0$ and take the difference to estimate the ATE. In our notation, we abbreviate the observation unit as $\bZ = (\bX,A,Y)$. For a (potentially) random function $f(\bZ)$, we denote $\MP_n \{f(\bZ)\}$ as its average over the samples, and $\MP\{f(\bZ)\} = \int f(\bz) d\MP(\bz)$ as the expectation where only the randomness of $\bZ$ is considered (so $f$ is conditioned on when it is random). Finally, we let $\| f\|^2 = \int f^2(\bz) d\MP(\bz)$ denote the squared $L_2$-norm. 

Next, we define two nuisance functions on which the estimated ATE depends, but that are not of direct interest themselves:
\begin{align*}
	& \pi(\bX) = \MP(A = 1 \mid \bX),  \quad \mu_a(\bX) = \ME(Y \mid A = a, \bX). 
\end{align*}
We refer to $\pi(\bX)$ and $\mu_a(\bX)$ as the propensity score and the outcome model, respectively. Estimation of these nuisance functions is a standard regression problem, for which any parametric or nonparametric method could be used. Next, we can define a ``plug-in'' estimator for $\psi_1$ based on the nuisance functions. Specifically, let $\psi_1$ be estimated as the average of an estimated outcome model: $\widehat{\psi}_1 = \MP_n\left[\hat{\mu}_1(\bX)\right]$, where $\hat{\mu}_1(\bX)$ is a regression estimator for $\mu_1$. The regression estimator for $\hat{\mu}_1(\bX)$ may be based on a flexible ML method relative to more restrictive parametric methods to avoid model misspecification. 

However, without further adjustments, this plug-in estimator for $\widehat{\psi}_1$ will typically inherit any first-order smoothing bias present in $\hat{\mu}_1(\bX)$ \citep{kennedy2016semiparametric, kennedy2022semiparametric}. Formally, we can characterize the conditional bias given the data used to estimate $\hat{\mu}_1$ as
\begin{equation}\label{eq:plugin-bias}
    \ME[\hat{\psi}_1 - \psi_1]= \ME\left[\hat{\mu}_1(\bX)-\mu_1(\bX)\right],
\end{equation}
In general, the bias term above would be of the same order as the error in estimating $\mu_1(X)$, which, in flexible, nonparametric models, is slower than $n^{-1/2}$.  We refer to the plug-in style estimator $\hat{\psi}_1 = \MP_n\{\hat{\mu}_1(\bX)\}$ as having first-order bias. An inverse-probability-weighted estimator such as $\MP_n\{ AY / \widehat\pi(X)\}$ would similarly have first order bias and inherit the slow rate of convergence of $\widehat\pi(X)$. Alternatively, one can find a function of the data, which we denote as $\varphi_1(\bZ)$, such that estimating $\psi_1$ as the average value of $\varphi_1$  will correct for this first-order bias and have smaller second-order bias. 
% Moreover, additional adjustments are needed to ensure that the estimator of $\psi_a$ does not inherit the convergence rate for estimating $\pi(\bx)$ and $\mu_a(\bx)$, which is slower than the parametric rate $n^{-1/2}$ for many flexible nonparametric and machine learning methods. 

For parameters like $\psi_a$, semiparametric efficiency theory offers a principled way to derive adjustments for these flaws. The core idea is to derive a first-order, functional Taylor expansion (von-Mises expansion) of the form:
\begin{align*}
    \psi(\hat{\mathbb{P}})-\psi(\mathbb{P})=-\int \phi(\bz,\hat{\mathbb{P}}) d \mathbb{P}(\bz)+R_2(\hat{\mathbb{P}}, \mathbb{P})
\end{align*}
for a mean-zero function; with $\phi(\bz; \MP)$ termed \textit{influence function}, and $R_2(\widehat\MP, \MP)$ is a second-order remainder term . This expansion motivates a correction to the plug-in estimator $\psi(\widehat\MP)$ leading to a new estimator $\widehat\psi =  \psi(\hat{\mathbb{P}}) + \MP_n\{ \phi(\bZ, \widehat\MP)\}$. In the case of the parameter $\psi_1$, the corrected estimator is known as the doubly-robust (augmented inverse-probability-weighted) estimator:
\begin{align*}
    \widehat{\psi}_1^{dr} = \MP_n \left[ \frac{A}{\widehat\pi(\bX)}\{Y-\hat{\mu}_1(\bX)\} + \hat{\mu}_1(\bX) \right] \equiv \MP_n\{\widehat\varphi_1(\bZ)\},
\end{align*}
because the influence function of $\psi_1$ is
\begin{align*}
    \phi_1(\bZ) \equiv \phi_1(\bZ; \MP) = \frac{A}{\pi(\bX)}\{Y - \mu_1(\bX)\} + \mu_1(\bX) - \psi_1(\MP)
\end{align*}
and $\varphi_1$ is the un-centered influence function \citep{robins1994estimation, lunceford2004stratification, kang2007demystifying}. 

The phrase \textit{doubly-robust} is motivated by the fact that $\widehat\psi_1^{dr}$ is consistent if either $\widehat\pi$ or $\widehat\mu_1$ (but not necessarily both) is consistent. When $\widehat\pi$ and $\widehat\mu_1$ are modeled sufficiently flexibly, one may expect that both functions are consistently estimated. Even if this is the case, one key advantage of $\widehat\psi_1^{dr}$ over other estimators not based on the influence function is that the error $\widehat\psi_1^{dr} -\psi_1$ will involve the \textit{product of errors} (in $L_2$) $\| \widehat\pi - \pi \| \| \widehat\mu_1 - \mu_1\|$, a central-limit-theorem (CLT) term and another term that is negligible under mild conditions. This can be seen from the decomposition:
\begin{align*}
    \widehat{\psi}_1^{dr} - \psi_1 & = \MP_n\{\hat{\varphi}_1(\bZ)\} - \MP\{\varphi_1(\bZ)\} \\
    & = (\MP_n - \MP) \{\varphi_1(\bZ)\} + \MP\{\widehat{\varphi}_1(\bZ) - \varphi_1(\bZ)\} + (\MP_n - \MP)\{\widehat{\varphi}_1(\bZ) - \varphi_1(\bZ)\}.
\end{align*}
The first term is the CLT term that is asymptotically normally distributed (when scaled by $\sqrt{n}$) with mean zero and variance $\ME\{\phi_1^2(\bZ)\}$. The second term evaluates to
\begin{align*}
    \MP\{\widehat\varphi_1(\bZ) - \varphi_1(\bZ)\} = \int \left\{\frac{\pi(\bx)}{\widehat\pi(\bx)} - 1 \right\}\{\mu_1(\bx) - \widehat\mu_1(\bx)\} d\MP(x),
\end{align*}
which implies that $\left| \MP\{\widehat\varphi_1(\bZ) - \varphi_1(\bZ)\} \right| \lesssim \|\widehat\pi - \pi\| \|\widehat\mu_1 - \mu_1\|$
by the Cauchy-Schwarz inequality as long as the estimated propensity score is bounded away from zero. This is crucial because it allows for a parametric rate of convergence for $\widehat\psi_1^{dr}$ even if flexible machine learning models are used for the nuisance functions as long as the product of their error rates is $o_\MP(n^{-1/2})$. This requirement can hold under structural assumptions, such as sparsity or smoothness. 

Finally, the third term is an empirical process term that is asymptotically negligible, i.e., $o_\MP(n^{-1/2})$, if $\|\widehat\varphi_1 - \varphi_1\| = o_\MP(1)$ and either 1) the function class where $\pi$, $\mu_1$, and their estimators reside is sufficiently regular (Donsker) or 2) the estimators $\widehat\pi$ and $\widehat{\mu}_1$ are computed on a separate, independent sample. The latter requirement can always be enforced in practice by 1) splitting the sample in $K$ different folds, 2) train the estimators in all but one fold and compute the estimator based on the remaining fold, and 3) averaging the resulting $K$ estimators of $\psi_1$ to compute the final estimator. We use this sample splitting approach in this paper. 

This general framework for constructing estimators based on the target's influence function has been referred to as targeted learning \citep{van2011targeted} or doubly robust machine learning (DRML) \citep{chernozhukov2018double,hernan2020causal}. See \citet{kennedy2022semiparametric} for a recent review. In this framework, DR methods are combined with ML estimation and sample-splitting to a construct a doubly robust estimator that is less sensitive to nuisance estimation errors when ML methods are used to estimate the nuisance functions. This, in turn, allows for $\sqrt{n}$-consistent inference even when ML methods are used to estimate the nuisance functions at slower than $\sqrt{n}$-rates. Critically, the DRML framwork is agnostic to the type of ML method used to estimate the nuisance functions and, in a particular sense, yields optimal estimators if no other assumption on the data generating process is introduced \citep{balakrishnan2023fundamental, bonvini2023drinference}. We summarize the DRML estimation procedure for $\psi_1$ in Algorithm \ref{alg:ate-estimation}.
\begin{algorithm} 
\label{alg:ate-estimation}
Input: iid data: $\bZ^n$, number of data folds: $K$. 
\begin{enumerate}
    \item Divide the sample $\bZ^n$ into $K$ folds of size approximately $n / K$. Let $I_k$ denote all units in split $k$ with $|I_k| = n_k$ and $I_k^c$ all units but those in split $k$. 
    \item Using only units in $I^c_k$, train $\hat{\pi}_{-k}(\cdot)$ and $\hat{\mu}_{1,-k}(\cdot)$ using suitable machine learning methods.
    \item Compute an estimate of $\psi_1$ on the test split $I_k$:
    \begin{align*}
    \widehat\psi_{1,k} = \frac{1}{n_k} \sum_{i \in I_k} \left[\frac{A_i}{\hat{\pi}_{-k}(\bX_i)} \{Y_i - \hat{\mu}_{1,-k}(\bX_i)\} + \hat{\mu}_{1,-k}(\bX_i)\right]
    \end{align*}
    \item Repeat steps 2-3 for each $k$ and set the final estimate to be
    \begin{align*}
    \widehat{\psi}_{1} = \frac{1}{K} \sum_{k = 1}^K \widehat\psi_{1,k}
    \end{align*}
    As an estimate of the variance, one can use
    \begin{align*}
    \widehat\sigma^2 = \frac{1}{K} \sum_{i = 1}^K \frac{1}{n_k} \sum_{i \in I_k} \left[\frac{A_i}{\hat{\pi}_{-k}(\bX_i)} \{Y_i - \hat{\mu}_{1,-k}(\bX_i)\} + \hat{\mu}_{1,-k}(\bX_i) - \widehat{\psi}_{1,k}\right]^2
    \end{align*}
    \item Report a Wald-type $1-\alpha$ confidence interval:
    \begin{align*}
    \widehat\psi_1 \pm z_{1-\alpha/2} \frac{\widehat\sigma}{\sqrt{n}}
    \end{align*}
\end{enumerate}
\end{algorithm}
%% Move algorithm to appendix.

\subsection{Application to LS for Cholecystitis}
\begin{figure}[htbp]
  \centering
    \includegraphics[scale=0.40]{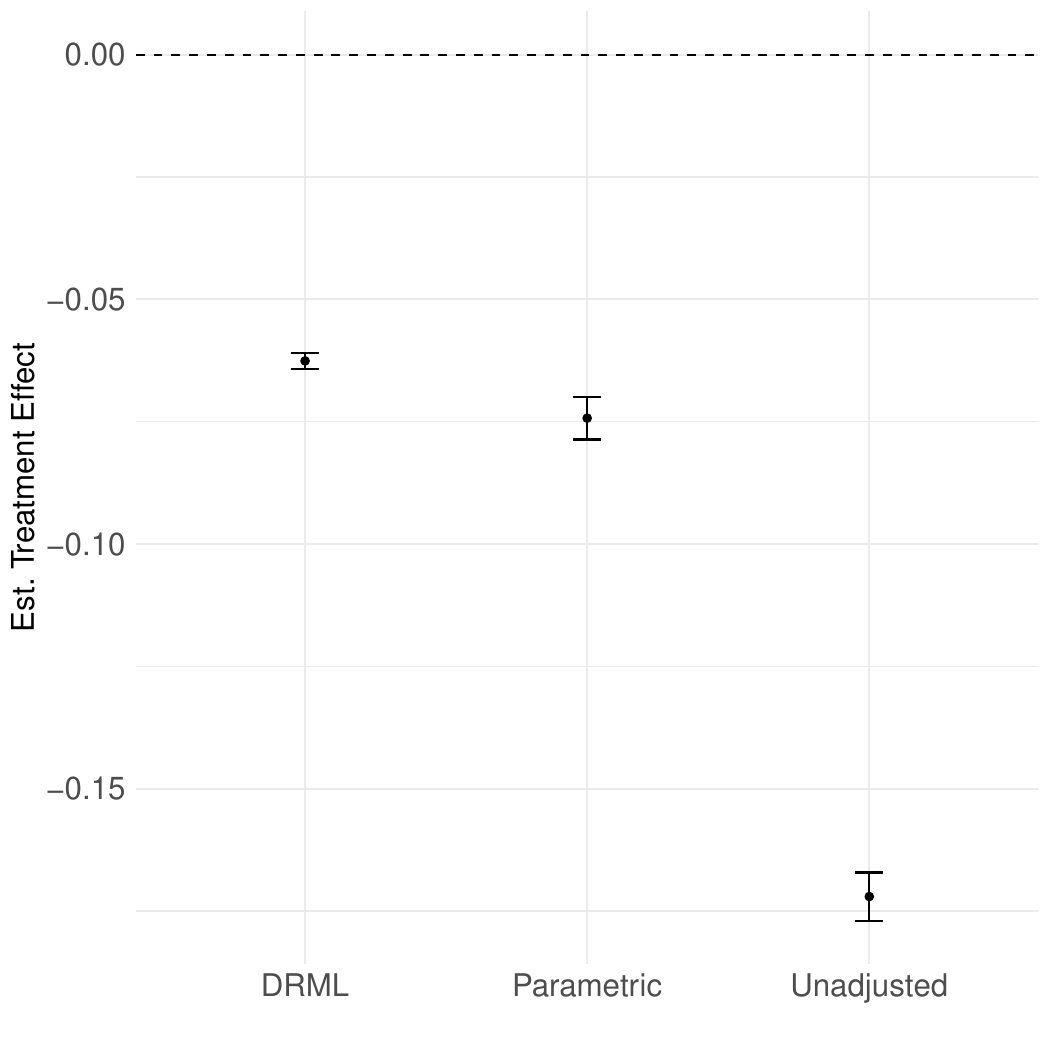}
    \caption{Estimates for the effect of LS on adverse events for patients with cholecystitis.}
  \label{fig:outs}
\end{figure}

In this section, we apply Algorithm \ref{alg:ate-estimation} to estimate the effect of LS on adverse events. To compute our estimator, we use ten-fold crossfitting and employ an ensemble of learners including random forests, the lasso, and boosted trees. We refer to the estimate as the DRML one, which we contrast with estimates from two other approaches, namely  1) the unadjusted effect of LS on adverse events and 2) the adjusted effect where the adjustment is based solely on parametric models. Figure~\ref{fig:outs} reports the three different estimates with corresponding confidence intervals. Notably, the unadjusted estimate shows a very large benefit of LS versus OS: the risk of an adverse event is 17\% lower for patients who underwent LS. However, this estimate may reflect the fact that LS patients are generally healthier than OS patients, and could overstate the magnitude of the LS effect.

Adjustment for baseline confounders reduces the magnitude of the estimate by more than half. Using DRML methods further reduces the magnitude of the estimated effect. However, the pattern across the two estimates is consistent. For patients that had LS for cholecystitis, the risk of an adverse event is 6 to 7\% lower. Given the large sample sizes, our estimates are also quite precise. Next, we explore whether there is any effect heterogeneity.

\section{Analysis of Treatment Effect Heterogeneity}

In this section, we assume that the analyst designates a subset of the covariates in $\bX$ as effect modifiers and the goal is to estimate the causal effects within levels of these covariates. For example, if we designate age as an effect modifier, we might expect the effect of LS to vary with age. We use $\bV \in \mathbb{R}^d $ to denote the subset of variables in $\bX$ that are effect modifiers. The variables in $\bV $ are selected \emph{a priori} based on clinical knowledge. When selected \emph{a priori}, $\bV$ is usually low-dimensional (fewer than five covariates are included). Later, we will conduct an exploratory analysis in Section \ref{sec:exploratory} to identify additional covariates in $\bX$ that may be effect modifiers. In this case, the set of effect modifiers may be high-dimensional. 

A common way to capture effect heterogeneity with respect to effect modifiers $\bV$ is through the conditional average treatment effect (CATE), defined as:
\[
\tau_v(\bv) = \ME[Y^1 - Y^0\mid\bV=\bv] .
\]
Under Assumption \ref{assumption:main} (consistency, positivity and ignorability), this estimand is identified as
\begin{align}\label{eq:cate}
\tau_v(\bv) = \ME[ \mu_1(\bX) - \mu_0(\bX)\mid\bV=\bv].
\end{align}
Notice that this is essentially the same identification formula as for the ATE (Eq. \ref{eq:ate}), except that the difference in outcome models $\mu_1(\bX) - \mu_0(\bX)$ is regressed on $\bV$ instead of being averaged with respect to the distribution of $\bX$. 

\subsection{Doubly-Robust Estimators of Conditional Average Treatment Effects}
\label{sec:ITE}
In this work, we estimate the CATE parameter as identified in Eq. \ref{eq:cate} using the DR-Learner framework developed in \citet{kennedy2020towards}, though many other approaches have been proposed, see e.g. \citet{kunzel2019metalearners}\footnote{The learners proposed in \cite{kunzel2019metalearners} were designed for applications where the propensity score is known---typically in randomized experiments. As such, they are inappropriate in our application because LS is not randomly assigned.}, \cite{foster2023orthogonal, morzywolek2023general} and references therein. Despite the similarities between the expression for the ATE \eqref{eq:ate} and the CATE \eqref{eq:cate}, the methods needed to efficiently estimate these two parameters are quite different. In fact, the CATE does not possess an influence function in nonparametric models whenever at least one component of $\bV$ is continuous. Given this, the approach described in Section \ref{sec:dr-ate} is not directly applicable to the CATE. However, recent work has shown that it is possible to leverage the theory developed for the ATE to obtain efficient estimators of the CATE. 

The core element of the DR-learner approach to CATE estimation is the regression of the un-centered influence function $\varphi^{\text{cate}}(\bZ)$ onto $\bV$, where
\begin{align*}
 \varphi^{\text{cate}}(\bZ; \pi, \mu_0, \mu_1)  \equiv  \varphi(\bZ) = {\mu}_1(\bX)-{\mu}_0(\bX)+\frac{\{A-{\pi}(\bX)\}\left\{Y-{\mu}_A(\bX)\right\}}{{\pi}(\bX)(1-{\pi}(\bX))}
\end{align*}
Notice that $\varphi^{\text{cate}}(\bZ)$ is the influence function for the ATE parameter ($\varphi^{\text{cate}}(\bZ) = \varphi_1(\bZ) - \varphi_0(\bZ)$ using the notation of Section \ref{sec:dr-ate}). When the goal is to estimate the ATE, one takes the average of an estimate of $\varphi^{\text{cate}}(\bZ)$, whereas, when the target is the CATE, the DR-Learner procedure prescribes regressing an estimate of $\varphi^{\text{cate}}(\bZ)$ onto $\bV$. In this light, $\varphi^{\text{cate}}(\bZ)$ takes the role of a \textit{pseudo-outcome} in the second-stage regression onto $\bV$. Just like when estimating the ATE, it is critical to estimate the nuisance functions, $\pi$, $\mu_1$ and $\mu_0$, via cross-fitting. We summarize the DR-Learner estimator of the CATE in Algorithm \ref{alg:cate-estimation}. 
\begin{algorithm} 
\label{alg:cate-estimation}
Input: iid data: $\bZ^n$, number of data folds: $K$. 
\begin{enumerate}
    \item Divide the sample $\bZ^n$ into $K$ folds of size approximately $n / K$. Let $I_k$ denote all units in split $k$ with $|I_k| = n_k$ and $I_k^c$ all units but those in split $k$. 
    \item Using only units in $I^c_k$, train $\hat{\pi}_{-k}(\cdot)$, $\hat{\mu}_{1,-k}(\cdot)$, and $\hat{\mu}_{0,-k}(\cdot)$ using suitable machine learning methods.
    \item Construct the estimated pseudo-outcome $\varphi^{\text{cate}}$ on $I_k$: for $i \in I_k$
    \[
    \widehat{\varphi}^{\text{cate}}(\bZ_i;\hat{\pi}_{-k}, \hat{\mu}_{0,-k}, \hat{\mu}_{1,-k}) = \widehat{\mu}_{1,-k}(\bX_i) - \widehat{\mu}_{0,-k}(\bX_i) + \frac{\{A_i - \widehat\pi_{-k}(\bX_i)\}\{Y_i - \widehat\mu_{A,-k}(\bX_i)\}}{\widehat\pi_{-k}(\bX_i)(1 - \widehat\pi_{-k}(\bX_i))}
    \]
    and regress $\widehat{\varphi}^{\text{cate}}(\bZ_i;\hat{\pi}_{-k}, \hat{\mu}_{0,-k}, \hat{\mu}_{1,-k}) $ on $\bV_i$ in the set $I_k$. Let $\hat{\tau}_{v,k}(\bv)$ be the estimator obtained. 
    \item Repeat steps 2-3 for each $k$ and set the final estimate to be
    \begin{align*}
    \hat{\tau}_{v} (\bv) = \frac{1}{K} \sum_{k = 1}^K \hat{\tau}_{v,k} (\bv)
    \end{align*}
\end{enumerate}
\end{algorithm}
As shown in \cite{kennedy2020towards}, the estimator described in Algorithm \ref{alg:cate-estimation} enjoys several appealing properties. First, consistency is achieved as long as either $\widehat\pi$ or both $\widehat\mu_1$ and $\widehat\mu_0$ are consistent. This double-robustness property is akin to that of the doubly-robust estimator of the ATE described in Algorithm \ref{alg:ate-estimation}. Second, under stability conditions on the second-stage regression, the bias due to the estimation of the nuisance functions involves only a product of errors of the form  $\|\widehat{\pi} - \pi\| \sum_{a \in \{0,1\}}\|\widehat{\mu}_a - \mu_a\|$. This property too is shared with the estimator of the ATE and it is crucial to attain fast convergence rates for estimating the CATE. Third, treatment effect heterogeneity can be estimated by CATEs evaluated at the full covariate vector $\bX$ or at a proper subset of effect modifiers $\bV \subset \bX$ by simply regressing $\widehat\varphi^{\text{cate}}(\bZ_i, \widehat\pi_{-k}, \widehat\mu_{0, k}, \widehat\mu_{1, -k})$ on either $\bX$ or $\bV$. In other words, Algorithm \ref{alg:cate-estimation} does not need any modifications whether $\bV$ or $\bX$ are used to define the CATE. This is in contrast with another popular approach to CATE estimation known as the R-Learner \citep{nie2021quasi}, where quite different methods may be required for $\bV \neq \bX$ specific tasks. 
\subsection{Assessing Whether There Exists Heterogeneity in the Effect of LS}

In this section, we investigate whether the effect of LS varies from patient to patient by estimating CATEs via the DR-Learner algorithm \ref{alg:cate-estimation} . For this analysis, we first need to identify key effect modifiers. As we noted earlier, based on clinical expertise, we identified sepsis status, age, and baseline risk of an adverse event as key effect modifiers. As such, we hypothesize that the effect of LS might be larger or smaller depending on whether patients have sepsis, are younger or have a lower baseline risk of an adverse event. In our analysis, we found that age and risk were highly correlated. Given this, we regressed the risk scores onto a smooth function of age within each level of sepsis status using the \texttt{gam} function from the \texttt{mgcv} package in \texttt{R}. We take the residuals from this model as a measure of residual risk that is not due to age and sepsis.

\begin{figure}[htbp]
  \centering
  \begin{subfigure}[t]{0.42\textwidth}
    \includegraphics[width=\textwidth]{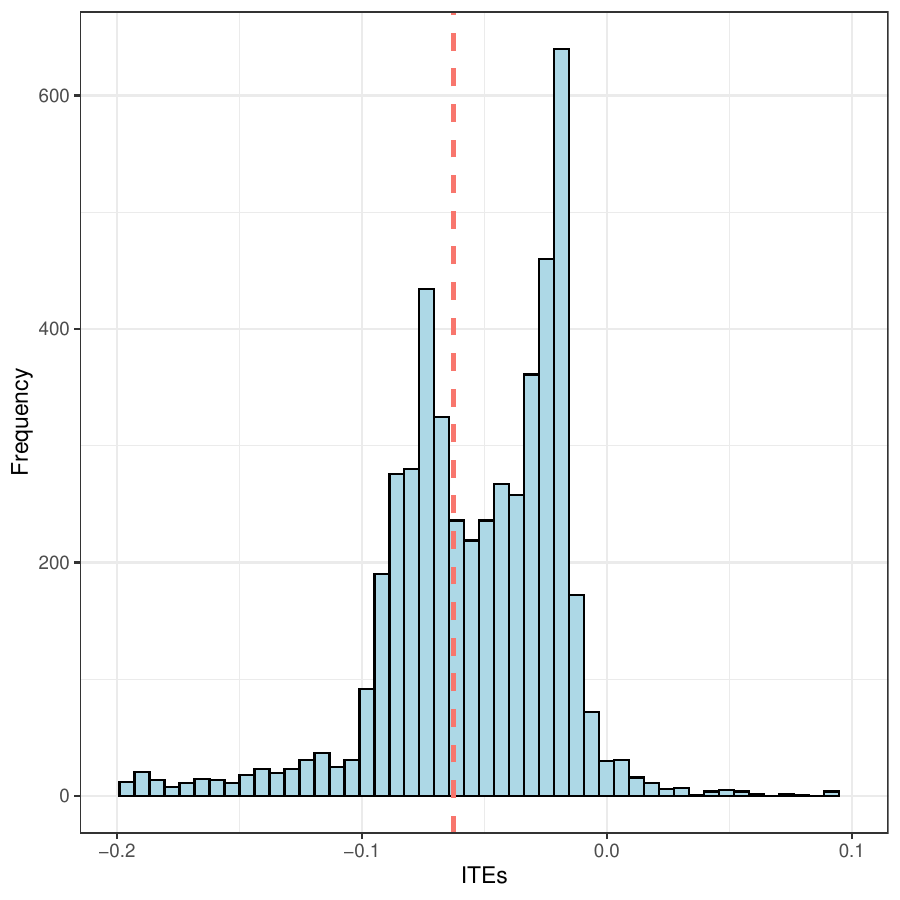}
    \caption{$E(Y^1 - Y^0 \mid V_i)$}
    \label{fig:cate-v}
  \end{subfigure}
  \begin{subfigure}[t]{0.42\textwidth}
    \includegraphics[width=\textwidth]{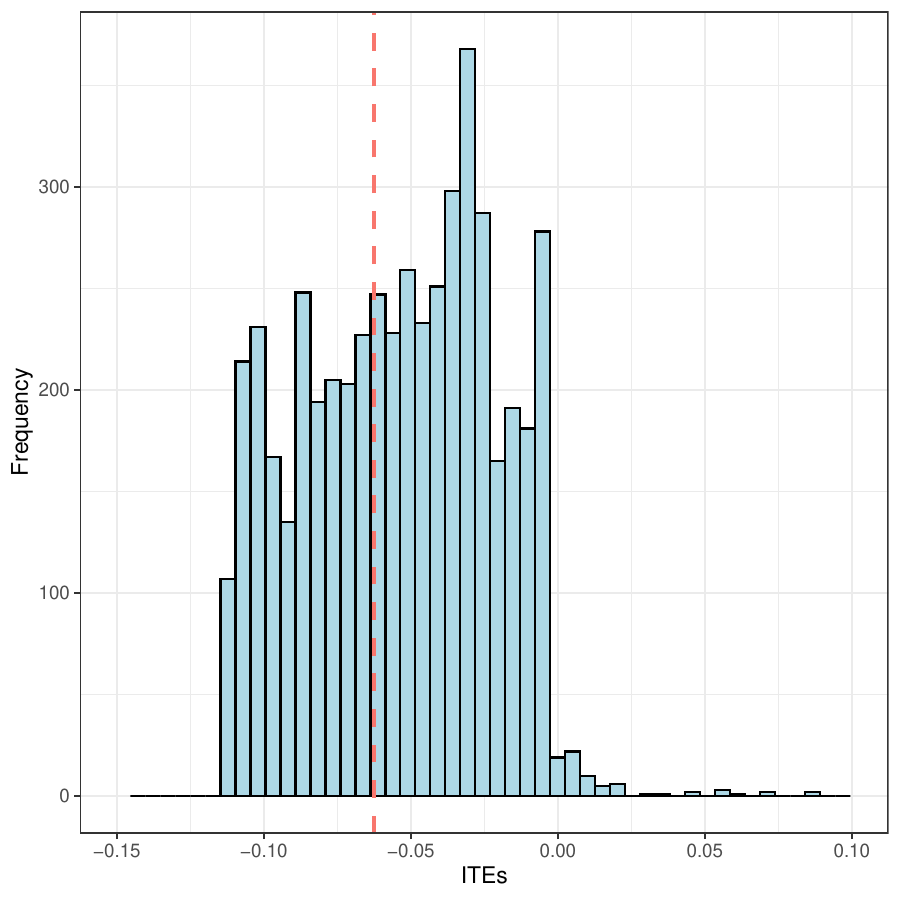}
    \caption{$E(Y^1 - Y^0 \mid X_i)$}
    \label{fig:cate-x}
  \end{subfigure}
  \caption{Distribution of individual treatment effects for the effect of LS on adverse events conditional on $V$ and $X$. The vertical dashed line demarcates the average treatment effect point estimate.}
  \label{fig:ites}
\end{figure}

To estimate the nuisance functions entering Algorithm \ref{alg:cate-estimation}, we rely on an ensemble of linear and generalized linear models, regression trees, splines, and a random forest, fitted using the SuperLearner \texttt{R} package \citep{van2007super}. We rely on 10-fold cross-fitting and do a single regression of the estimated pseudo-outcomes onto the relevant covariates in $\bV$ or $\bX$. The second-stage regression of the pseudo-outcome on age, risk and sepsis is fitted by least-squares on additive basis splines predictors with interactions between the bases and the indicator for sepsis. The order of the bases is computed by leave-one-out-cross-validation (LOOCV). In Figure~\ref{fig:ites}, we plot the distribution of the CATE evaluated at $\bV$ and $\bX$. When the CATE is evaluated at the full vector of measured covariates $\bX$, it is often referred to as an Individual Treatment Effect (ITE).

Both of these distributions show considerable variation in the effect of LS on an adverse event. Moreover, their support is mostly in a negative interval of the real-line, indicating that, for most patients, LS reduces the probability of an adverse event. There seem to be relatively few patients that are harmed by LS as evident from the light right-tails of these distributions.  When we condition on $\bX$, we observe that there is more variation in the distributions of the ITEs\footnote{This is not surprising because, for $Y = Y^1 - Y^0$, $X = (X_1, X_2)$ and $V = X_1$:
\begin{align*}
   \text{Var}\{\ME(Y \mid X_1, X_2)\} - \text{Var}\{\ME(Y \mid X_1)\} & = \ME\{\ME(Y \mid X_1, X_2)^2\} - \ME\{\ME(Y \mid X_1)^2\} \\
   & = \ME[\ME\{\ME(Y \mid X_1, X_2)^2 \mid X_1\}] - \ME\{\ME(Y \mid X_1)^2\} \\
   & \geq \ME\{\ME(Y \mid X_1)^2\} - \ME\{\ME(Y \mid X_1)^2\} = 0
\end{align*}}
Figure \ref{fig:ites}, while useful to investigate whether there is treatment effect heterogeneity, provides little evidence about which effect modifiers are most responsible for the variation observed. In the next section, we study several ways to improve the interpretation and visualization of the estimated CATEs.

\section{Methods for the Interpretation of CATEs in the Meta-Learner Framework}
\label{sec:interpret-CATE}
In more traditional parametric model-based approaches to CATE estimation, quantities of interest can be derived from estimated parameters as marginal effects (e.g. with a linear outcome model the estimated coefficients multiplying the effect modifiers may be used for interpretation). In a machine-learning-based framework, where nonparametric methods are used, interpretation may be difficult due to the possible black-box nature of the algorithms involved.

When all the variables in $\bV$ are discrete, estimation of the CATE can be done by simply fitting a saturated model. That is, one  computes the subgroup-specific ATE by taking the average of the estimated $\varphi^{\text{cate}}(\bZ)$ within the relevant subgroup. For example, in our analysis, one of the effect modifiers is an indicator for sepsis. The estimated CATE evaluated at just this variable would simply be the average of the ITEs for patients with and without sepsis. However, if $\bV$ contains at least one continuous covariate, such as age or the measure of risk, some form of smoothing instead of simple averaging is necessary. In this case, additional modeling will generally be needed for the final estimate of the CATE. Next, we propose three different ways to visualize and interpret the effects of continuous effect modifiers. Our goal is to strike a good balance between flexibility, in order to avoid model misspecification, and interpretability for better visualization and straightforward inference.

\subsection{Univariate CATE Curves}
\label{sec:individual-cate}

The first strategy is to estimate the CATE for each effect modifier and visualize the modifier's effect as a univariate function. Mathematically, we estimate $\tau_j(v_j) = \ME[Y^1 - Y^0\mid V_j = v_j]$ for each $j \in \{1,\dots,d\}$. By the tower property of conditional expectation, 
\[
\tau_j(v_j) = \ME\{\varphi^{\text{cate}}(\bZ;\pi,\mu_0,\mu_1)|V_j =  v_j\}.
\]
Estimating $\tau_j(v_j)$ can be reduced to performing a univariate regression of $\varphi^{cate}(\bZ, \widehat\pi, \widehat\mu_0, \widehat\mu_1)$ onto $V_j$. This can be done following Algorithm \ref{alg:cate-estimation}, with the second-stage regression taking the form of a univariate nonparametric regression. Because the second-stage regression is univariate, we propose using relatively simple nonparametric methods for which inferential procedures are well-understood. For example, one could use local polynomial regression or nonparametric least-squares methods. In our analysis, we estimate $\tau_j(v_j)$ via local linear regression. 

Just like for any standard nonparametric regression, estimating $\tau_j(v_j)$ when $v_j$ is continuous requires a trade-off between bias and variance. In particular, it is well-known that traditional confidence bands obtained under optimal smoothing are centered around a smoothed version of the true regression function, as opposed to the regression function itself. This is referred as the ``bias problem'' in nonparametric smoothing (see e.g. Section 5.7 in \cite{wasserman2006all}). There are at least two ways to address this bias problem. One way is to ignore the bias and acknowledge that the inference is for the smoothed version of $\tau_j(v_j)$ instead of the function $\tau_j(v_j)$ itself. We report estimates and confidence bands covering the smoothed parameter with a desired probability. 

A second approach is to estimate the smoothing bias and subtract it off from the original estimator. Building upon \cite{calonico2018effect}, \citet{takatsu2022debiased} 
 has recently studied this approach for estimating the dose-response function. We adapt their methods to estimate $\tau_j(v_j)$, which can be done by simply replacing their pseudo-outcome with $\varphi^{\text{cate}}(\bZ)$. The inferential guarantees of their methods would then hold in our case as well, under suitably modified regularity conditions. 
One potential drawback of debiasing is that it requires the estimation of the second (or higher) derivative of the target regression function, which is typically harder to estimate than the regression function itself. However, one does not need an optimal estimator of such higher order derivative for the debiasing approach to work. As long as the target function possesses some additional smoothness not exploited by the original estimator without debiasing, the confidence band based on the debiasing approach has the nominal coverage probability asymptotically \citep{takatsu2022debiased}. 
An alternative approach to debiasing would be to undersmooth the original estimator; that is, to choose a localizing bandwidth of smaller order than the optimal one minimizing the mean-square-error in order to render the smoothing bias asymptotically negligible (at the expense of an increase in the variance and a suboptimal rate). Because we are not aware of practical guidance on how to choose the right amount of undersmoothing, we do not pursue this approach in this paper. 

To better describe the estimator of $\tau_j(v_j)$ that we implement, we focus on estimating $\tau_1(v_1)$ as an example. In the first stage, following Algorithm \ref{alg:cate-estimation}, we use suitable machine learning methods to estimate the nuisance functions $\pi, \mu_0, \mu_1$ and construct the pseudo-outcome: $ \widehat{\varphi}^{\text{cate}}(\bZ_i) = \varphi^{\text{cate}}(\bZ_i; \hat{\pi}, \hat{\mu}_0, \hat{\mu}_1)$. In the second stage, we estimate $\tau_1$ at a fixed point $v_1^0$ by regressing $\widehat{\varphi}^{\text{cate}}(\bZ_i)$ onto $V_1$ with a local linear estimator. Specifically, let
\[
\bg_{h, v_1^0, j}(v_1):=\left[1,\frac{v_1-v_1^0}{h}, \ldots, \left(\frac{v_1-v_1^0}{h}\right)^j\right]^\top
\]
be the $j$-th order polynomial basis around $v_1^0$, where $h=h_n$ is the bandwidth parameter shrinking to zero as $n$ goes to infinity. Let $K_{h,v_1^0}(v_1) = \frac{1}{h}K\left( \frac{v_1-v_1^0}{h}\right)$ be a kernel function, e.g. $\one(|v_1 - v_0| \leq h) / (2h)$. The local linear estimator of $\tau_1(v_1^0)$ is 
\[
\hat{\tau}_1^{\text{LL}}(v_1^0)={\be}_1^{\top}\hat{\bbb}_{h,1}(v_1^0),
\] where $\be_j$ is a $(j+1)$-dimensional vector with $j$-th component being 1 and others being 0, while $\hat{\bbb}_{h,1}(v_1^0)$ solves the following weighted least square problem:
\[
\hat{\boldsymbol{\beta}}_{h,1}(v_1^0)=\underset{\boldsymbol{\beta} \in \mathbb{R}^2}{\arg \min } \mathbb{P}_n\left[K_{h, v_1^0}(V_1)\left\{\widehat{\varphi}^{\text{cate}}(\bZ ; \hat{\pi}, \hat{\mu}_0, \hat{\mu}_1)-\mathbf{g}_{h, v_1^0,1}(V_1)^{\mathrm{T}} \boldsymbol{\beta}\right\}^2\right].
\]
The closed-form solution is 
\[
\hat{\tau}_1^{\text{LL}}(v_1^0)=\be_1^{\top}\widehat{\bD}_{h, v_1^0, 1}^{-1} \MP_n[\bg_{h,v_1^0, 1}(V_1)K_{h,v_1^0}(V_1) \widehat{\varphi}^{\text{cate}}(\bZ)],
\]
where $\widehat{\bD}_{ h, v_1^0, j}=\MP_n\left[\bg_{h, v_1^0, j}(V_1) \bg_{h, v_1^0, j}(V_1)^\top K_{h, v_1^0} (V_1)\right]$. When $h$ is chosen by cross-validation, confidence bands will be centered around a smoothed version of CATE $\tilde{\tau}_1(v_1)=\tau_1(v_1)+\frac{1}{2}h^2c_2 \tau_1^{(2)}(v_1)$, where $\tau_1^{(2)}$ is the second-order derivative of $\tau_1$ and $c_2=\int u^2 K(u) d u$. In our analysis, we also implement the debiasing approach described in \cite{takatsu2022debiased}, which corrects for the bias by explicitly estimating the second-order derivative with a third-order local polynomial. Let $b=b_n$ be the bandwidth parameter; we solve the following minimization problem:
\[
\hat{\boldsymbol{\beta}}_{b,3}(v_1^0)=\underset{\boldsymbol{\beta} \in \mathbb{R}^4}{\arg \min } \mathbb{P}_n\left[K_{b, v_1^0}(V_1)\left\{\widehat{\varphi}^{\text{cate}}(\bZ ; \hat{\pi}, \hat{\mu}_0, \hat{\mu}_1)-\mathbf{g}_{b, v_1^0,3}(V_1)^{\top} \boldsymbol{\beta}\right\}^2\right].
\]
The closed-form solution of $\hat{\tau}_1^{(2)} = 2b^{-2}\be_{3}^{\top}\hat{\boldsymbol{\beta}}_{b,3}(v_1^0)$ is
\[
\hat{\tau}_1^{(2)} (v_1^0) = 2 b^{-2} \be_3^\top \widehat{\mathbf{D}}_{ b, v_1^0, 3}^{-1} \mathbb{P}_n\left[\bg_{b, v_1^0, 3} (V_1)K_{b, v_1^0}(V_1) \widehat{\varphi}^{\text{cate}}(\bZ)\right].
\]
Thus, the debiased estimator is
\[
\hat{\tau}_1^{\text{de}}(v_1^0) =\mathbb{P}_n\left[\hat{\Gamma}_{h, b, v_{1}^0}(V_1) \widehat{\varphi}^{\text{cate}}(\bZ) \right],
\]
where
\[
\hat{\Gamma}_{h, b, v_{1}^0}(v_1)=\be_1^T \widehat{\mathbf{D}}_{h, v_1^0, 1}^{-1} \bg_{h, v_1^0, 1}\left(v_1\right) K_{h, v_1^0}\left(v_1\right)-  c_2 h^2 b^{-2} \be_3^T  \widehat{\mathbf{D}}_{ b, v_1^0, 3}^{-1} \bg_{b, v_1^0, 3}\left(v_1\right) K_{b, v_1^0}\left(v_1\right)
\]
Pointwise and uniform (in $v_1$) confidence bands can be constructed by following the methods described in \cite{takatsu2022debiased} based on simulating from a Gaussian process with appropriate covariance function. More details can be found in Appendix \ref{appendix:individual-cate-band}, including a characterization of the influence function of the localized parameter $\tau_{h, b}\left(v_1^0\right)=\mathbb{E}\left[\Gamma_{ h, b, v_1^0}\left(V_1\right) \tau_1\left(V_1\right)\right]$ ($\Gamma_{ h, b, v_1^0}\left(V_1\right)$ is the population version of $\hat{\Gamma}_{h, b, v_{1}^0}(V_1)$), which plays a critical role in the construction of uniform confidence bands. 

The function $\tau_j(v_j)$ represents a meaningful way to summarize and understand how the treatment effect varies with modifier $V_j$. However, it may mask cases where the heterogeneity with respect to $V_j$ is entirely due to other variables. For instance, $\tau_v(\bV)$, the CATE as a function of all effect modifiers $\bV$, may only depend on $v_k$ in the form of $\tau_v(\bV) = g(V_k)$ for some function $g$ and modifier $V_k \neq V_j$. Then the direct effect of $V_j$ on the treatment should thus be homogeneous. Yet $\tau_j(v_j) = \ME[\tau_v(\bV)\mid V_j=v_j] = \ME[g(V_k)\mid V_j=v_j]$, which may not necessarily be constant in $v_j$ if $V_j$ is not independent of $V_k$. Hence $\tau_j(v_j)$ may not be an ideal summary of heterogeneous effects of $V_j$ in this setting. In the next sections, we outline methods that can better capture the effect heterogeneity due to $V_j$ when all the other effect modifiers are taken into account.

\subsection{Additive CATE Curves}
\label{sec:gam}
In nonparametric regression, assuming that the regression function can be decomposed into a sum of univariate functions, one for each covariate, often strikes a good balance between generalizability and interpretability. Such additive modeling assumptions can be easily incorporated into the estimation of the  CATEs as well. After all, Algorithm \ref{alg:cate-estimation} would reduce to a standard nonparametric regression problem had the nuisance functions be known. Thus, in this section, we estimate the CATE function by assuming the following additive structure: $\tau_v(\bv) = \alpha + \sum_{j=1}^d h_j(v_j)$. For identification purposes, it is assumed that $\ME[h_j(V_j)] = 0$, for $j \in \{1,\dots, d \}$. When all the marginal effects $h_j$ can be estimated accurately, this approach correctly detects that the CATE is constant in $V_j$ whenever $\tau_v(\bV) = g(V_k)$ for $V_j \neq V_j$. 

For estimation, one can regress ${\varphi}^{\text{cate}}(\bZ; \widehat\pi,\widehat\mu_0,\widehat\mu_1)$ on $\bV$ using any generalized additive model (GAM) estimation procedure. In this paper, we fit the GAM model by nonparametric least-squares so that we can rely on the theoretical results from \cite{semenova2021debiased} to construct valid uniform confidence bands around each $h_j(v_j)$. By the additive assumption on $\tau_v(\bv)$, we are allowed to choose univariate basis functions $\bb_j(v_j)\in \mathbb{R}^m$  for each effect modifier $V_j$ and no interactions. We then estimate the best linear approximation of $\tau_v$ in the linear span of $\{ \bb_j(v_j) , 1\leq j \leq d\}$ by least-squares:
\[
\min_{\alpha \in \mathbb{R}, \bbb_j \in \mathbb{R}^m} \sum_{i=1}^n \left(\widehat{\varphi}^{\text{cate}}(\bZ_i;\hat{\pi},\hat{\mu}_0,\hat{\mu}_1) - \alpha - \sum_{j=1}^d \bb_j^{\top}(v_{ij})\bbb_j\right)^2.
\]
In many applications, the number of effect modifiers is usually small. It is therefore possible to select the dimension of basis $m$ to be moderately large to guarantee small approximation error while maintaining $(md + 1) \ll n$ so that the least-square estimator is well-defined. The estimate of the marginal effect $h_j(v_j)$ is $\bb_j^{\top}(v_j)\hat{\bbb}_j - \frac{1}{n}\sum_{i=1}^n \bb_j^{\top}(v_{ij})\hat{\bbb}_j$. Under the additional regularity conditions in \cite{semenova2021debiased}, we can construct confidence bands for $h_j$ by applying their method. We fix $\bv_{-j}$ at $\bv_{-j}^0$ while letting $v_j$ vary across a grid and apply the multiplier bootstrap, which yields valid confidence bands for 
\[
\bb_j^{\top}(v_j)\bbb_j + \sum_{k \neq j} \bb_k^{\top}(v_k^0)\bbb_k.
\]
After centering it at $0$ we obtain a confidence band for the projection of $\tau_j$ onto the linear span of $\bb_j$.

Our proposed GAM modeling strategy allows us to jointly consider the effect of each effect modifier and hence potentially decouple the indirect effects due to correlation between covariates. At the same time, the additive structure facilitates estimation (e.g., no multivariate basis is needed) and interpretation (i.e., we can understand the heterogeneous effects by visualizing each marginal effect $h_j$). There are, however, two potential drawbacks: some effect modifiers may function interactively so the CATE $\tau_v$ may not be expressed as summation of univariate functions. In this case, using a GAM may oversimplify the problem and suffer from model misspecification. Second, in terms of estimation, our method relies on the number of effect modifiers being relatively small so that $(md + 1) \ll n$ holds to guarantee the consistency of estimates. Regularization is necessary when the number of effect modifiers $d$ and number of basis $m$ are both large so as to avoid overfitting. In this more challenging setting, we may use other estimation procedures to fit the GAM, including backfitting algorithms \citep{hastie1987generalized, hastie2009elements} and a reduced rank smoothing approach (penalized maximum likelihood estimation) \citep{wood2004stable, wood2011fast, wood2016smoothing}. We leave the study of CATE estimation and inference using these approaches as an avenue for future work.

\subsection{Partial Dependence CATE Curves}
\label{sec:partial-dependence}

Inspired by \cite{zhao2021causal}, in this section, we propose a method to simultaneously address potential violations of the additivity assumption in GAMs, while retaining the advantages of estimating univariate regression functions. This approach is based on the concept of \textit{partial dependence}, which, in the machine learning literature, is typically used to assess the impact of a single covariate on the predictions from a black-box model. Formally, we propose estimating
\[
\theta_j\left(v_j\right)=\int \tau_v(\bv) d \mathbb{P}\left(\bv_{-j}\right)=\mathbb{E}\left\{\mathbb{E}\left[\mu_1(\bX)-\mu_0(\bX) \mid V_j=v_j, \bV_{-j}\right]\right\}
\]
as the partial dependence function. Note that to compute $V_j$-specific CATEs, $\tau_j(v_j) = \int \tau_v(\bv) d \MP (\bv_{-j}\mid v_j)$, we take expectation over $\bV_{-j}$ with respect to the conditional distribution of $\bV_{-j}$ given $V_j$. However, to compute $\theta_j(v_j)$, the expectation is taken with respect to the marginal distribution of $\bV_{-j}$.

One advantage of the partial dependence function is that, if additivity holds, it can correctly recover the $V_j$-specific component. To see this, suppose that $\tau_v(\bv) = \alpha + g_j(v_j)+g_{-j}(\bv_{-j})$ for $\ME[g_j(V_j)] = 0$, then $\theta_j(v_j)$ equals $g_j(v_j)$ up to an additive constant, so that $g_j$ can be recovered by estimating $\theta_j(v_j)$ together with a centralization step. If additivity does not hold, $\theta_j(v_j)$ can still be a useful measure of effect heteronegeity because it captures how the effect varies with $V_j$ \textit{on average}, where, at each evaluating point $v_j$, the average is computed including units who are unlikely to be observed with $V_j = v_j$. Importantly, if the heterogeneity is entirely due to variables other than $V_j$, then $\theta_j(v_j)$ would be a constant function. We view the partial dependence function as a way to capture effect heterogeneity that is complementary with respect to the other measures discussed in previous sections. 

To estimate $\theta_j(v_j)$ and perform inference efficiently, we follow Algorithm \ref{alg:cate-estimation} but we use a different pseudo-outcome in the second stage regression. We also carry out standard nonparametric inference and the debiased inference from \cite{takatsu2022debiased} exactly the same way as discussed in Section \ref{sec:individual-cate}, modulo the different pseudo-outcome construction. Specifically, let $\eta = \left(\pi, \mu_0, \mu_1, f_{j|-j}\right)$ be the nuisance functions, where $f_{j|-j}$ is the conditional density of $V_j$ given $\bV_{-j}$. In agreement with the DR-Learner framework exemplified in Algorithm \ref{alg:cate-estimation}, we propose estimating $\theta_j(v_j)$ by regressing the following pseudo-outcome onto $V_j$:
\[
\begin{aligned}
\varphi_{j}^{\text{pd}}\left(\bZ; \eta \right)= &\, \left[\frac{\{A-\pi(\mathbf{X})\}\left\{Y-\mu_A(\mathbf{X})\right\}}{\pi(\mathbf{X})\{1-\pi(\mathbf{X})\}}+\tau_x(\mathbf{X})-\mathbb{E}\{\tau_x(\mathbf{X}) \mid \mathbf{V}\}\right] \frac{f_j(V_j)}{f_{j|-j}\left(V_j \mid \mathbf{V}_{-j}\right)} + \theta_j(V_j)\\
=&\, \left[\frac{\{A-\pi(\mathbf{X})\}\left\{Y-\mu_A(\mathbf{X})\right\}}{\pi(\mathbf{X})\{1-\pi(\mathbf{X})\}}+\tau_x(\mathbf{X})-\mathbb{E}\{\tau_x(\mathbf{X}) \mid \mathbf{V}\}\right] \frac{\int_{\mathcal{V}_{-j}}f_{j|-j}\left(V_j|\bv_{-j}\right) d\MP (\bv_{-j})}{f_{j|-j}\left(V_j \mid \mathbf{V}_{-j}\right)} \\
&\, +  \int_{\mathcal{V}_{-j}} \ME[\tau_x(\bX)\mid V_j, \bV_{-j} = \bv_{-j}]  d \mathbb{P}\left(\bv_{-j}\right),
\end{aligned}
\]
where $\tau_x(\bX) = \mu_1(\bX) - \mu_0(\bX)$ is the conditional average treatment effects in terms of all covariates $\bX$. As shown in Appendix \ref{appendix:partial_dependence}, the pseudo-outcome $\varphi_{j}^{\text{pd}}\left(\bZ; \eta \right)$ possesses a double-robustness property in that regressing it on $V_j$ yields a consistent estimator of $\theta_j(v_j)$ as long as either $(\widehat{\mu}_0, \widehat{\mu}_1)$ or $(\widehat{\pi}, \widehat{f}_{j|-j})$ are consistent estimators. As for inference, one can apply standard nonparametric procedures, e.g. based on nonparametric series or local polynomial regression. As in Section \ref{sec:individual-cate}, we conduct both standard and debiased inference using a second-stage local linear regression of an estimate of $\varphi_{j}^{\text{pd}}\left(\bZ; \eta \right)$ onto $V_j$. Pointwise and uniform confidence bands can be constructed based on the influence function of the smoothed parameter $\theta_{h, b,j}\left(v_j^0\right)=\mathbb{E}\left[\Gamma_{ h, b, v_j^0}\left(V_j\right) \theta_j(V_j)\right]$, where $\Gamma_{h, b, v_j^0}$ is defined in Section \ref{sec:individual-cate}. More details, including the characterization of the influence function, can be found in Appendix \ref{appendix:partial_dependence}. 

\subsection{Comparison of the Three Approaches to Effect Heterogeneity Estimation}

We conclude this section with a general discussion on the three methods introduced above to visualize and interpret the role of effect modifiers. Perhaps the most straightforward way to visualize effect heterogeneity is to estimate univariate, or $V_j$-specific, CATEs as outlined in Section \ref{sec:individual-cate}. This method yields easily interpretable results: the estimated function is the average treatment effect among the relevant subgroup population. However, this approach cannot distinguish between effect heterogeneity directly attributable to $V_j$ or ``indirect'' heterogeneity due to the correlation between $V_j$ and the other effect modifiers. This phenomenon is similar to falsely including irrelevant, yet collinear variables in variable selection problems \citep{fan2010selective}. To address this potential issue while retaining straightforward visualization of effect heterogeneity, in Section \ref{sec:gam}, we propose an additive structure for $\tau_v$ to allow for modeling the impact of all effect modifiers simultaneously. Each component of the additive model can be plotted together with confidence bands. In some applications, however, an additive structure might be hard to justify. To overcome this challenge, in Section \ref{sec:partial-dependence}, we propose using a partial dependence function, motivated by the usage of this function in interpreting black-box machine learning models \citep{zhao2021causal}. The partial dependence function also helps filter out indirect effects and recover the additive structure of the CATE in case it holds in practice. Although it does not appear to be widely used in studies of heterogeneous treatment effects, we believe this parameter has appealing properties for clinical applications with multiple effect modifiers.

\subsection{Measuring Heterogeneity in the Effect of LS}

In this section, we apply the methods from Section \ref{sec:interpret-CATE} to estimate the CATEs for the three effect modifiers in our study: age, risk, and sepsis status. For the two multi-valued effect modifiers, age and risk, we compare and contrast the CATEs based on univariate, additive, and partial dependence methods. The estimated ITEs computed following Algorithm \ref{alg:cate-estimation} are shown in Figure\ref{fig:cate-v}. For each method, the nuisance functions are estimated using the SuperLearner ensemble method \citep{van2007super} with linear models, generalize linear models, regression trees, splines and Random Forest libraries. When estimating the partial dependence function $\theta_j(v_j)$, we follow the approach taken in \cite{kennedy2017nonparametric} to estimate the conditional density; we model it semiparametrically by estimating the conditional mean and variance using a GAM while using a kernel density estimate for the density of the residuals.  When estimating $V_j$-specific CATEs or partial dependence functions, the local linear regression bandwidth is computed by leave-one-out cross-validation (LOOCV). Similarly, when estimating the GAM from Section \ref{sec:gam}, the number of basis elements is chosen by LOOCV.

\begin{figure}[htbp]
  \centering
    \begin{subfigure}[t]{0.6\textwidth}
    \includegraphics[width=\textwidth]{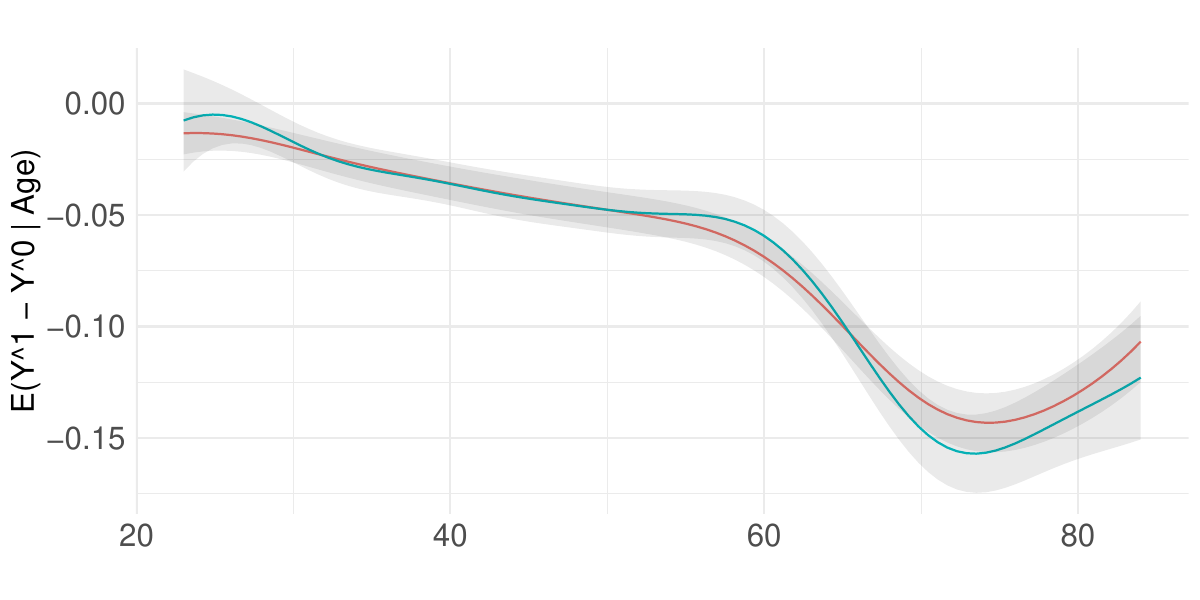}
    \caption{Univariate}
    \label{fig:age.1}
  \end{subfigure}
    \begin{subfigure}[t]{0.6\textwidth}
    \includegraphics[width=\textwidth]{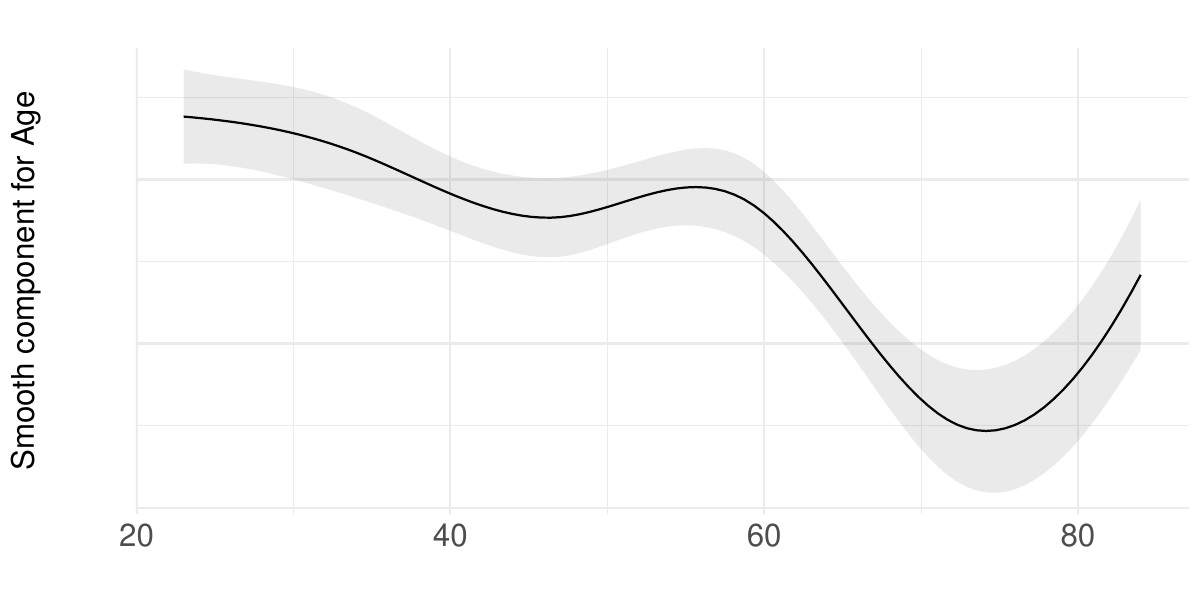}
    \caption{Additive}
    \label{fig:age.2}
  \end{subfigure}
  \begin{subfigure}[t]{0.6\textwidth}
    \includegraphics[width=\textwidth]{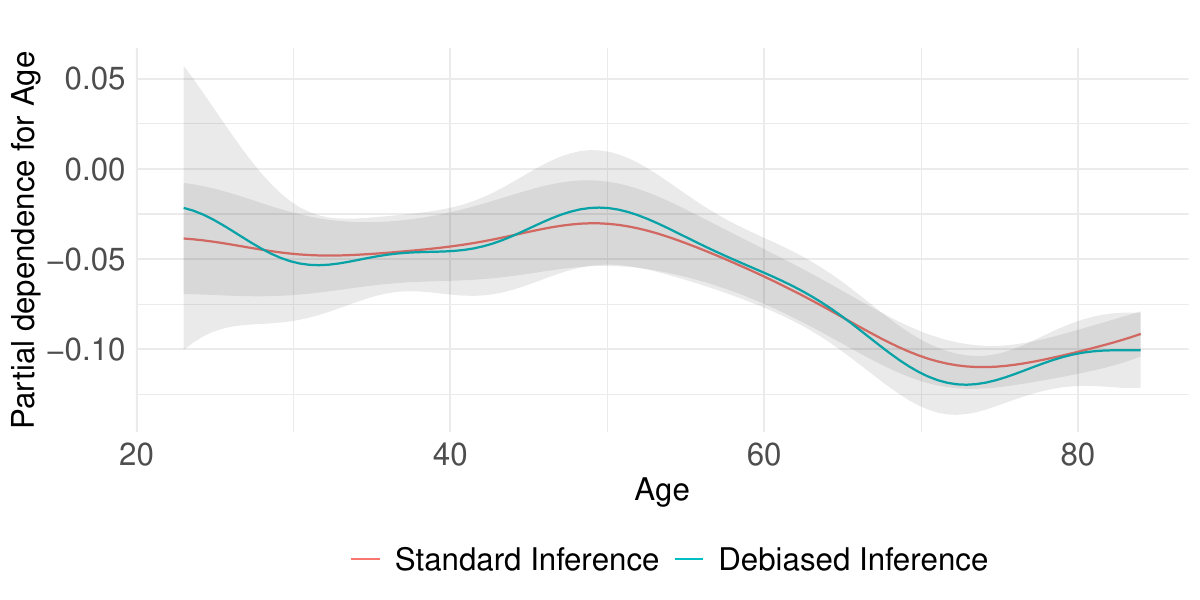}
    \caption{Partial Dependence}
    \label{fig:age.3}
  \end{subfigure}  
  \caption{CATE for Age effect of LS on Adverse Outcomees}
  \label{fig:cate.age}
\end{figure}
Figure~\ref{fig:cate.age} contains the estimates of effect heterogeneity based on age from the three estimation methods. The results in panel~\ref{fig:age.1} are based on the univariate, $V_j$-specific CATE method from Section \ref{sec:individual-cate}. We observe that, for older patients, LS reduces the risk of an adverse event.  The risk of an adverse event is notably lower for patients above the age of 60. However, for patients above the age of 70 the gains in risk reduction are somewhat reversed. The results in panel~\ref{fig:age.2} are based on the GAM modeling assumption from Section \ref{sec:gam}, while those in panel~\ref{fig:age.3} reflects our estimates of partial dependence function from Section \ref{sec:partial-dependence}. Overall, we find that the conclusions regarding effect heterogeneity based on age do not vary substantially with the methods used. That is, the additive and partial dependence methods produce results that are quite similar to the univariate method, which ignores potential correlation between age and risk. For the univariate, $V_j$-specific CATE and the partial dependence functions, we notice that the standard nonparametric inference and debiased inference are in substantial agreement. These results underscore the utility of using flexible methods to estimate treatment effect heterogeneity: being able to incorporate potential non-linearity in the effect estimates allowed us to uncover how LS produces the largest benefits for patients between the age of 60 and 70.
\begin{figure}[htbp]
  \centering
    \begin{subfigure}[t]{0.6\textwidth}
    \includegraphics[width=\textwidth]{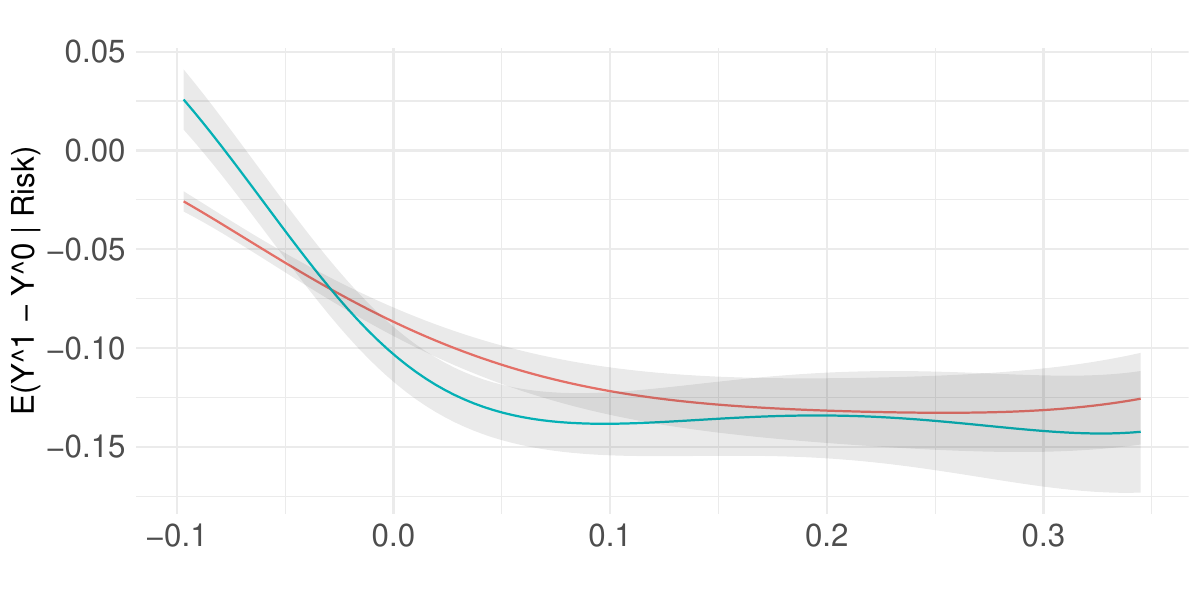}
    \caption{Univariate}
    \label{fig:risk.1}
  \end{subfigure}
    \begin{subfigure}[t]{0.6\textwidth}
    \includegraphics[width=\textwidth]{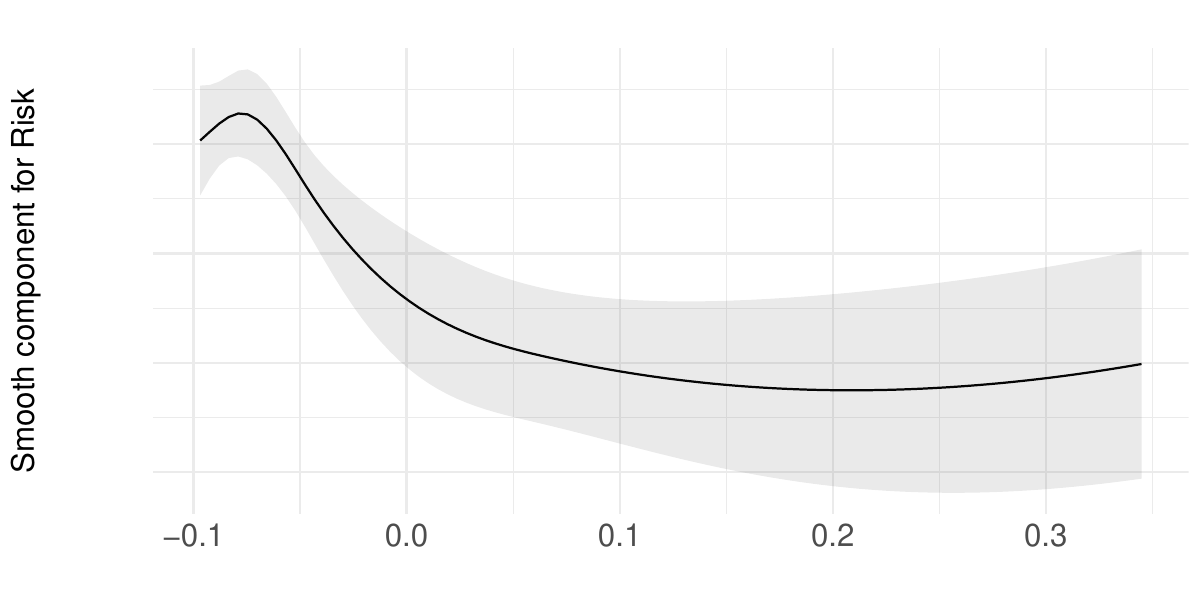}
    \caption{Additive}
    \label{fig:risk.2}
  \end{subfigure}
  \begin{subfigure}[t]{0.6\textwidth}
    \includegraphics[width=\textwidth]{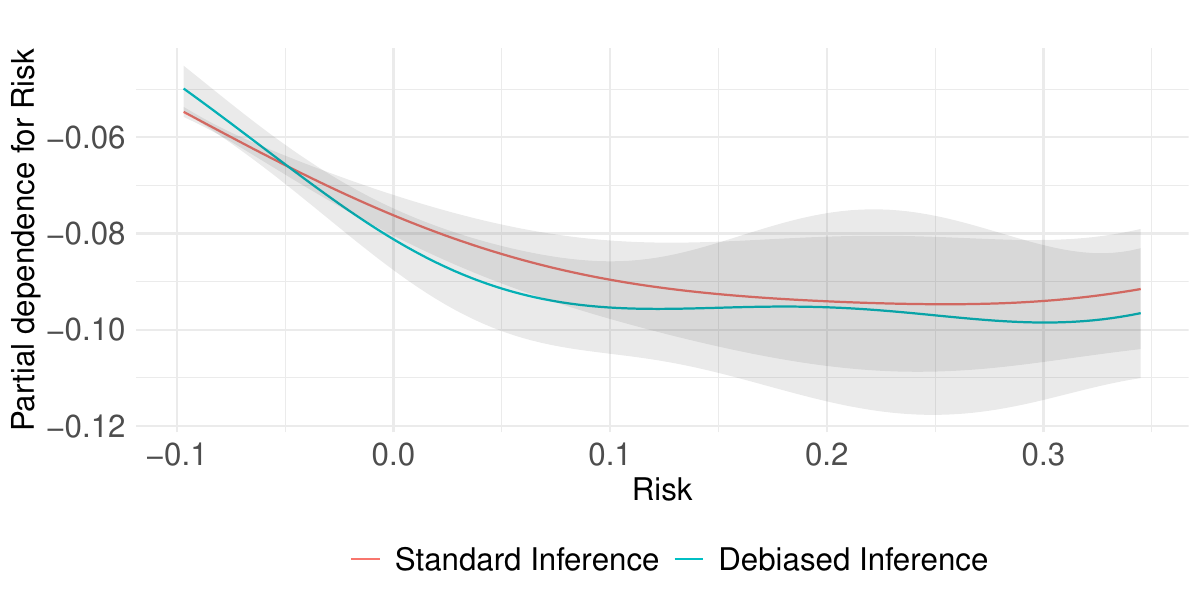}
    \caption{Partial Dependence}
    \label{fig:risk.3}
  \end{subfigure}  
  \caption{CATE for Risk effect of LS on Adverse Outcomees}
  \label{fig:cate.risk}
\end{figure}

Next, we consider treatment effect heterogeneity with respect to risk, which is the residual risk of an adverse outcome that is not explained by age and sepsis status. Figure~\ref{fig:cate.risk} reports the estimates based on the three methods from Section \ref{sec:interpret-CATE}. Similarly to the case of heterogeneity based on age, we find that the effect of LS on adverse outcomes does not vary linearly with risk. Across all three methods, we find a threshold effect: as risk increases, the benefit of LS increases until itstabilizes and is roughly constant for higher levels of risk. There are some slight differences between the univariate and partial dependence methods. When heterogeneity is measured by the partial dependence function, the range of the change in risk is smaller than when it is measured by the risk-specific, univariate CATE. In terms of inference, we find some differences between the standard inferential approach and the debiased one when heterogeneity is measured by a risk-specific CATE across low-to-mid values of risk (Figure \ref{fig:risk.1}).

To capture effect heterogeneity based on the indicator for sepsis, we simply marginalize the ITEs within units with and without sepsis. We find that the risk of an adverse event for non-septic patients is $-0.06$ (95\% CI:  $ -0.07 , -0.058 $).  This estimate is nearly identical to the unconditional estimate of the ATE. However, the risk of an adverse event for septic patients is  $-0.12$ (95\% CI: $  -0.15 , -0.10$). As such, sepsis also appears to be an important effect modifier. While both sets of patients benefit from LS relative to open surgery, the risk reduction due to LS is notably larger for septic patients.   

\section{Simulation Study}
\label{sec:simulation}

In this section, we assess the performance of the methods discussed above via a simulation study. We consider the following setting: $\bX = (W, \bV)$, where $W \sim N(0,1)$ is a baseline covariate, independent of $\bV$, and $ \bV = (V_1, V_2) \sim N(\mathbf{0}, \Sigma)$ are the effect modifiers. We set the components of $\Sigma$ as $\sigma_{11} = \sigma_{22}= 1, \sigma_{12} = \sigma_{21} = \rho$. Let the propensity score $\pi(\bX) = \MP(A=1\mid \bX) = \text{expit}(0.4W-0.2V_1-0.2V_2)$ so that given $\bX$, $A\sim \text{Bernoulli}(\pi(\bX))$. The outcome models for treated and control group are $\mu_1(\bX) = W + 1.5 V_1 - 0.5V_2$ and $\mu_0(\bX) = 0.5W + 0.5V_1 - 1.5V_2$, respectively. The final outcome is $Y = A\mu_1(\bX) + (1-A)\mu_0(\bX) + N(0,1)$. Under this data-generating process, the CATE function is
\[
\tau_x(\bX) = 0.5W + V_1 + V_2
\]
\[
\tau_v(V_1, V_2) = V_1 + V_2, \; (V_1,V_2) \sim N(\mathbf{0}, \Sigma).
\]
The univariate CATE and partial dependence functions are $\tau_1(V_1) = (1+\rho)V_1$ and $ \theta_1(V_1) = V_1$, respectively. Since the underlying CATE function has an additive structure, we expect GAM modeling discussed in Section \ref{sec:gam} to recover the partial dependence in $\theta_1(V_1)$. In our simulation study, we estimate the CATE using: 1) the univariate CATE estimation procedure from Section \ref{sec:individual-cate}, 2) the GAM modeling from Section \ref{sec:gam} and 3) the partial dependence estimation procedure from Section\ref{sec:partial-dependence}. The first procedure estimates $\tau_1(V_1)$, while the other two procedures estimate $\theta_1(V_1)$, which, unless $\rho = 0$, is not equal to $\tau_1(V_1)$. The nuisance functions $\pi, \mu_1, \mu_0$ are estimated using correctly specified parametric models while the conditional density $f_{1|-1}$ is estimated by a Gaussian distribution with fitted models for conditional mean and variance of $V_1$ given $V_2$. We use two-folds for sample splitting for all methods. In the study, we repeat each simulation scenario 500 times. We compute the root mean-squared error (RMSE) as
\[
\widehat{\text{RMSE}}_{\tau} = \int_{\mathcal{V}_1^*} \left[\frac{1}{M}\sum_{m=1}^M \left(\widehat{\tau}_{1}^m(v_1) - \tau_1(v_1)\right)^2 \right]^{1/2} d \MP(v_1),
\]
\[
\widehat{\text{RMSE}}_{\theta} = \int_{\mathcal{V}_1^*} \left[\frac{1}{M}\sum_{m=1}^M \left(\widehat{\theta}_{1}^m(v_1) - \theta_1(v_1)\right)^2 \right]^{1/2} d \MP(v_1),
\]
where $\mathcal{V}_1^*=[-2,2]$ denotes a trimmed support of $V_1$. In our simulation, we also include the oracle version of each DR-learner (where the true nuisance functions are used in constructing pseudo-outcomes) as a benchmark. 

\begin{figure}[htbp]
  \centering
    \begin{subfigure}[t]{0.8\textwidth}
    \includegraphics[width=\textwidth]{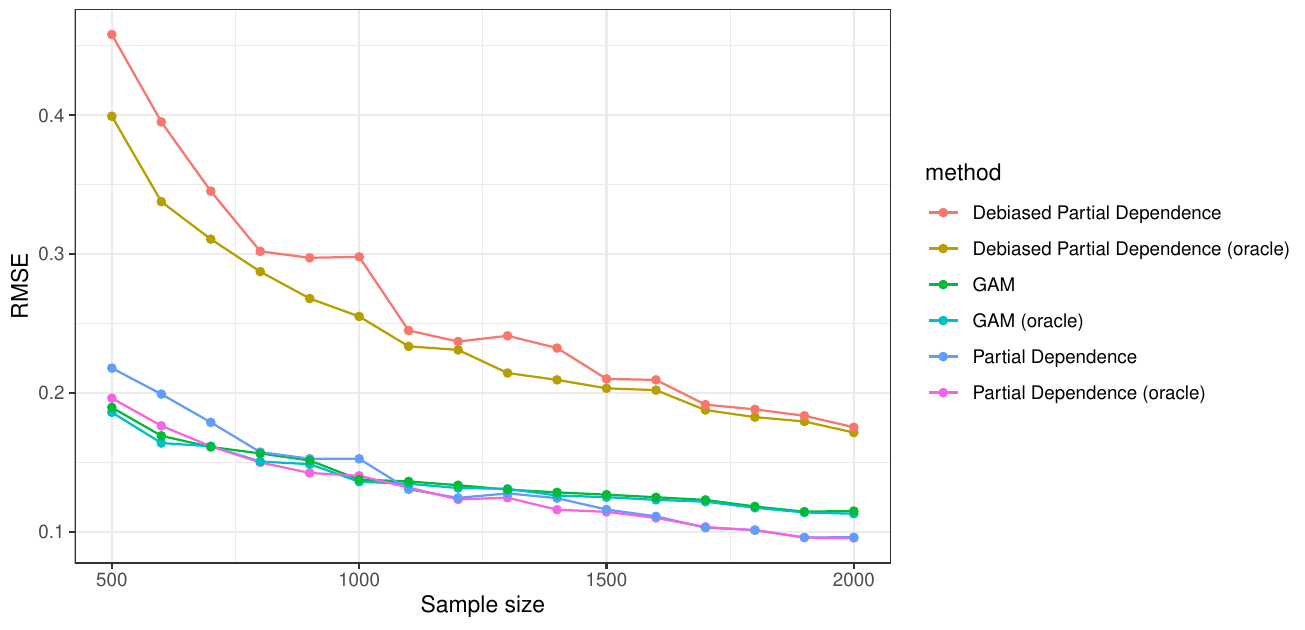}
    \caption{Scenario 1: Sample size}
    \label{fig:simu_pd1}
  \end{subfigure}
    \begin{subfigure}[t]{0.8\textwidth}
    \includegraphics[width=\textwidth]{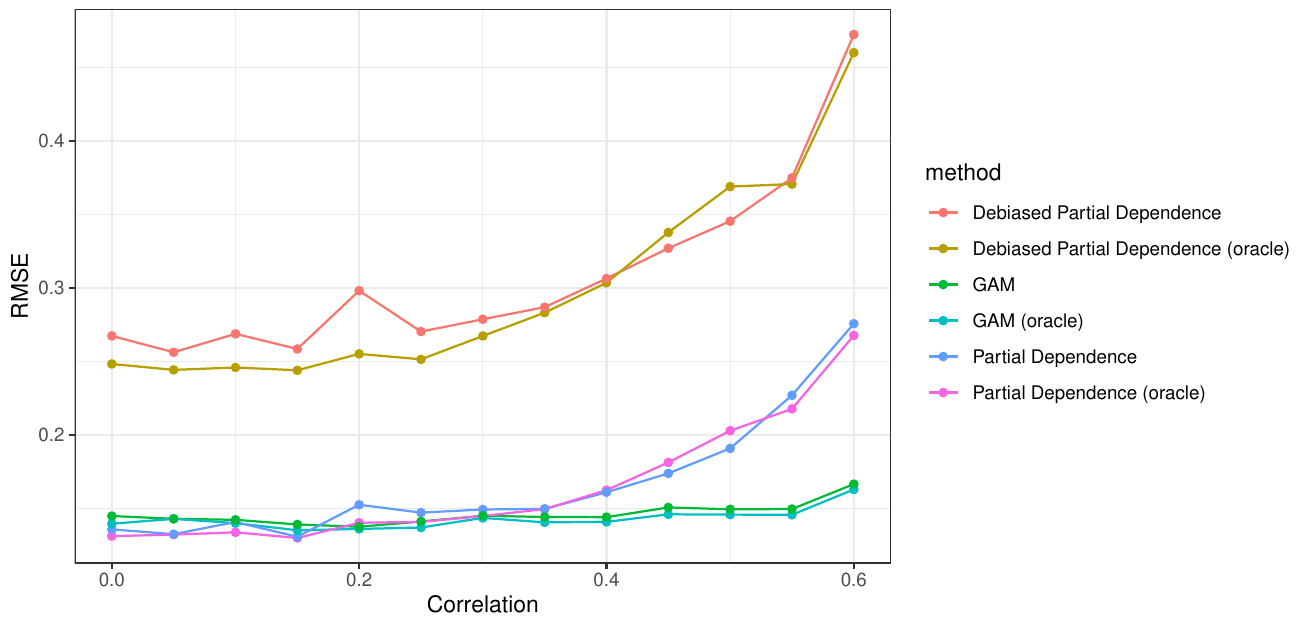}
    \caption{Scenario 2: Correlation $\rho$}
    \label{fig:simu_pd2}
  \end{subfigure}
  \caption{Simulation results for estimating partial dependence function $\theta_1$ under two scenarios.}
  \label{fig:simu_pd}
\end{figure}

We consider two scenarios. First, we fix the correlation coefficient $\rho=0.2$ and explore the relationship between sample size and RMSE. In this experiment, we vary the sample size from 500 to 2000 observations. The results for estimating the partial dependence function $\theta_1$ are summarized in Figure \ref{fig:simu_pd}. Results for univariate CATEs are presented Appendix \ref{appendix:addi-simu}. As expected, as the sample size increases, the estimation error of each method decreases. Overall the debiased inference method has larger estimated RMSE than the method that ``lives with the bias" and simply target a smoothed approprimation of the estimand. 
This result is consistent with the intuition that the reduction in bias from the debiasing approach may come at the cost of inflated variance (see discussion in Section 3 in \cite{takatsu2022debiased}). 
Notably, the GAM and partial dependence methods have nearly identical RMSE for smaller sample sizes, but for sample sizes larger than 1500, PD begins to outperform the GAM.

In the second scenario, we fix the sample size to $n=1000$ and vary the correlation $\rho$ between $V_1$ and $V_2$. The results from this study are presented in Figure \ref{fig:simu_pd2}. As expected, the methods to estimate partial dependence are all sensitive to the correlation $\rho$. As the correlation increases, it's harder to estimate the partial dependence function $\theta_1$. For the GAM modeling, this is similar to multicollinearity in the usual regression problem, where correlation between explanatory variables induces larger estimation error. For the partial dependence function, the correlation between $V_1$ and $V_2$ may affect how stable the estimate of the conditional density in $\varphi_j^{\text{pd}}$ is and hence influence the estimation error. Notably, GAM modeling does not appear to be as sensitive to the correlation between $V_1$ and $V_2$ as the estimators of the partial dependence functions are. In fact, it yields smaller RMSE than the approach targeting the partial dependence function. Just like in the first scenario, we observe that debiased inference comes at the price of an inflated RMSE. 
\section{Exploratory Analyses for Interpretation and Inference in the Meta-Learner Framework}
\label{sec:exploratory}

Thus far in our analysis, we have focused on flexibly estimating effect heterogeneity for three key effect modifiers that were identified \emph{a priori} based on clinical judgement. In this section, we focus on a exploratory analysis where we seek to identify a set of effect modifiers from the larger set of baseline covariates. More specifically, we seek to identify a subset of variables from $\bX$ that may be important effect modifiers of LS. To that end, we conduct an analysis where we estimate variable importance measures for effect modification following the approach described in \cite{hines2022variable}. In this analysis, we estimate a measure of treatment effect variable importance (TE-VIMP) for the full set of baseline covariates. Variables with large TE-VIMP values may be regarded as important effect modifiers.

Before computing the variable importance measures in our data analysis, we provide a brief review of the method developed by \cite{hines2022variable}. We redefine the notation slightly. As before, $\bX$ remains the full set of baseline covariates. Variables $\bV \subseteq \bX$ represent possible candidate effect modifiers. Note that $\bV$ may be specified as a single covariate in $\bX$ or a set of covariates. Let $\bX \backslash \bV$ denote the set of baseline covariates with the covariates in $\bV$ removed. Mathematically, for a set of covariates $\bV$, the TE-VIMP score is defined as
\[
\Psi_{v} = \frac{\Theta_{v}}{\operatorname{Var}\{\tau_x(\bX)\}} = 1 - \frac{\operatorname{Var}\{\ME[\tau_x(\bX)\mid \bX \backslash \bV]\}}{\operatorname{Var}\{\tau_x(\bX)\}},
\] 
and $\Theta_{{v}}$ is defined as
\[
\Theta_{v} := \ME[\operatorname{Var}\{ \tau_x(\bX)\mid \bX \backslash \bV \} = \operatorname{Var}\{\tau_x(\bX)\} - \operatorname{Var}\{\ME[\tau_x(\bX)\mid \bX \backslash \bV]\},
\]
where $\tau_{x\backslash v}(\bX) = \ME[\tau_x(\bX)\mid \bX \backslash \bV] $ is the CATE in terms of $\bX \backslash \bV$. The TE-VIMP score ranges from 0 to 1. It measures the amount of variation in $\tau_x(\bX)$ that could not be explained by $\bX \backslash \bV$, where a larger value indicates that the variables in $\bV$ explain a larger amount of the variation in effect heterogeneity. The TE-VIMP score is a nonparametric counterpart to the ANOVA statistics and the coefficient of determination $R^2$. We iterate over the variables in $\bX$ to estimate the TE-VIMP for each of the variables, although in principle one could compute the TE-VIMP measure for each subset of $\bX$ of interest.

\cite{hines2022variable} describes an influence-function based estimator of $\Psi_v$ that also employs a sample-splitting scheme in a way similar to our Algorithms \ref{alg:ate-estimation} and \ref{alg:cate-estimation} (see their Algorithms 1 and 2). In particular, they show that $\Theta_{v}$ can be efficiently estimated as:
\[
\widehat{\Theta}_{v} = \MP_n \left[\left( \widehat{\varphi}^{\text{cate}}(\bZ) - \widehat{\tau}_{x\backslash v}(\bX) \right)^2 - \left(  \widehat{\varphi}^{\text{cate}}(\bZ) - \widehat{\tau}_x(\bX)\right )^2 \right].
\]
and $\widehat{\Psi}_v = \widehat{\Theta}_v \backslash \widehat{\Theta}_x$, where $\widehat{\varphi}^{\text{cate}}, \widehat{\tau}_{x\backslash v}, \widehat{\tau}_x$ are constructed from a separate independent sample. Under suitable regularity conditions, \cite{hines2022variable} prove the $\sqrt{n}$-consistency and asymptotic normality of $\widehat{\Theta}_{v}$ and $\widehat{\Psi}_v$, from which one can derive Wald-type confidence intervals for $\Theta_v$. 

\subsection{Discovering Additional Potential Effect Modifiers of the Effect of LS}

\begin{figure}[htbp]
  \centering
    \includegraphics[scale=0.75]{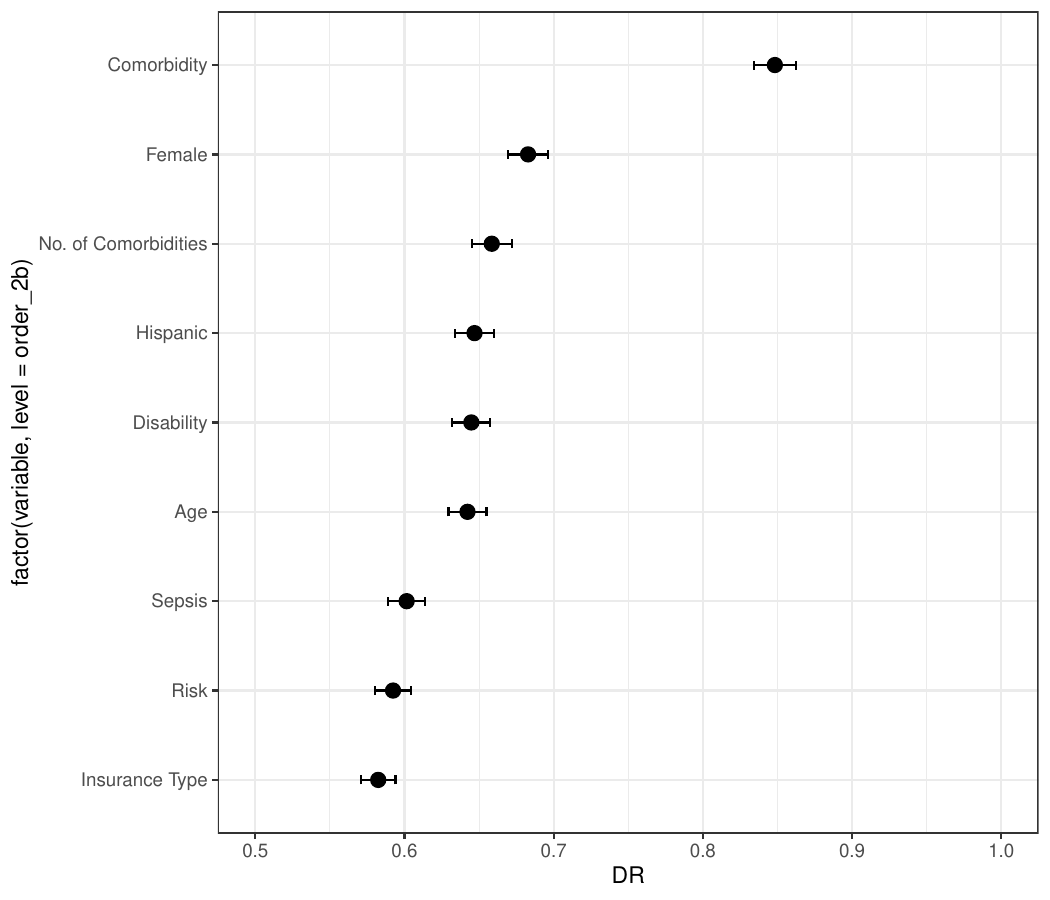}
    \caption{Estimates of TE-VIMP for Effect Modification of LS versus Open Surgery}
  \label{fig:vimp}
\end{figure}

We conclude our analysis of heterogeneity in the treatment effect of LS on adverse outcomes by computing the variable importance measure described in the previous section. To estimate $\Theta_v$, we use the same ensemble of learners that we used to fit the ITEs (Section \ref{sec:ITE}) and five-fold cross-fitting. In our data, there are 31 indicators for different types of comorbidities. As these measures are highly correlated, we report the importance of including versus excluding all of them simultaneously from the model.  Figure~\ref{fig:vimp} reports our estimates of TE-VIMP for each baseline covariate. A number of covariates have TE-VIMP scores that are essential zero and are omitted from the plot. Chief among these covariates with zero values are racial categories. We find that a variable indicating the presence of a comorbidity has by far the largest TE-VIMP at 0.85. The rest of the covariates have TE-VIMP values that range from just under 0.60 to 0.70.  Taking sampling variability into account, we find that the variables female, number of comorbidities, hispanic, disability status, and age have higher and overlapping TE-VIMP values. The variables sepsis status, risk of an adverse event, and insurance type have lower and overlapping TE-VIMP values. In general, this analysis reveals that a large fraction of the baseline covariates may be important effect modifiers for the LS treatment. It also suggests that, while there is strong evidence that LS is generally beneficial for patients, there is substantial variation in the extent to which patients benefit. 

\section{Conclusion}
In studies focused on the identification and estimation of treatment effects, the most common numerical summary is the ATE . Being an aggregate measure of treatment efficacy, the ATE may mask considerable effect variation between units. In our study, we used a large observational data source and modern statistical methods to 1) estimate whether patients that undergo Laparoscopic surgery instead of open surgery have fewer adverse events and 2) measure the extent to which the effect of LS varies with baseline covariates. To that end, we focused on two different methodological developments to improve our study.

First, we developed a framework for studies of heterogeneous treatment effects that allows for both confirmatory and exploratory analyses. We used data from randomized trials to benchmark and corroborate our estimate of the ATE. We then exploited the larger sample sizes and richer covariate sets in the observational data to investigate effect heterogeneity. We outlined two forms of analysis. The first analysis uses clinical expertise to identify a set of candidate effect modifier variables and then conducts a confirmatory test.  The secondary analysis uses the full set of baseline covariates to find the subset of variables that are important effect modifiers, i.e. that are responsible to large portions of the variation in the CATE function.

We developed new statistical tools to more easily interpret effect heterogeneity estimates. For continuous effect modifiers, we proposed three ways to summarize and visualize how treatment effects vary with a given effect modifier. The first approach is to simply consider univariate CATE functions; the second approach is to specify a general additive model so that each individual component can be estimated and visualized; the third approach is to compute the partial dependence function. The partial dependence function has the advantage of avoiding measuring heterogeneity not directly due to the effect modifier of interest, while preserving straightforward visualization and inference. All our estimators can efficiently incorporate modern machine learning methods. 

When the proposed methodology was applied to our data, we found that age, risk, and sepsis were all important modifiers of the LS effect. The variation in the effect of LS varied in a nonlinear fashion with age, and to a more limited extent, with risk. In our exploratory analysis, we found that the presence of a comorbidity was a particularly important effect modifier. Moreover, another eight covariates were identified as significant effect modifiers of the LS effect based on the variable importance measure considered. We caution that these exploratory results should be confirmed by additional studies. To conclude, the pattern that emerges from our data analysis is that LS is generally beneficial to patients relative to open surgery. However, there is considerable variation in its efficacy; the benefits are substantially larger for older, septic patients and those with a higher baseline risk of an adverse event.

\clearpage

\bibliographystyle{asa}
\bibliography{learner}

\appendix

\section{Construction of confidence bands for the local linear estimator in Section \ref{sec:individual-cate}\label{appendix:individual-cate-band}}

To construct uniform confidence band, we need to estimate the covariance of the estimator $\hat{\tau}_1^{\text{de}}$ evaluated at different points. The localized, smoothed parameter $\tau_{h, b}\left(v_1^0\right)=\mathbb{E}\left[\Gamma_{ h, b, v_1^0}\left(V_1\right) \tau_1\left(V_1\right)\right]$ is useful for this purpose, where
\[
{\Gamma}_{h, b, v_{1}^0}(v_1)=\be_1^\top {\mathbf{D}}_{h, v_1^0, 1}^{-1} \bg_{h, v_1^0, 1}\left(v_1\right) K_{h, v_1^0}\left(v_1\right)-  c_2 h^2 b^{-2} \be_3^\top {\mathbf{D}}_{ b, v_1^0, 3}^{-1} \bg_{b, v_1^0, 3}\left(v_1\right) K_{b, v_1^0}\left(v_1\right),
\]
and ${\bD}_{ h, v_1^0, j}=\ME\left[\bg_{h, v_1^0, j}(V_1) \bg_{h, v_1^0, j}(V_1)^\top K_{h, v_1^0} (V_1)\right]$ are the population versions of $\hat{\Gamma}_{h, b, v_{1}^0}(v_1)$ and $\widehat{\bD}_{ h, v_1^0, j}$, respectively.

The following theorem summarizes the efficient influence function of $\tau_{h, b}\left(v_1^0\right)$.

\begin{theorem}\label{thm:EIF-cate}
For each $h, b > 0$ and $v_1^0 \in \mathcal{V}_1$, consider a nonparametric model class consisting of $\MP$ such that under $\MP$, $\ME [Y^2] < \infty$. Then, the efficient influence function of $\tau_{h, b}\left(v_1^0\right)$ under this model class is
\[
\begin{aligned}
\phi_{ h, b, v_1^0}^{\text{cate}}(\bZ)= & \Gamma_{ h, b, v_1^0}\left(V_1\right) {\varphi}^{\text{cate}}(\bZ)-\gamma_{h, b, v_1^0}^{\text{cate}}\left(V_1\right), \text { where } \\
\gamma_{ h, b, v_1^0}^{\text{cate}}\left(V_1\right)= & \be_1^\top \mathbf{D}_{ h, v_1^0, 1}^{-1} K_{h, v_1^0}\left(V_1\right) \bg_{h, v_1^0, 1}\left(V_1\right) \bg_{h, v_1^0, 1}^\top\left(V_1\right) \mathbf{D}_{ h, v_1^0, 1}^{-1} \mathbb{E}\left[\bg_{h, v_{1}^0,1}\left(V_1\right) K_{h, v_1^0}\left(V_1\right)  \tau_1\left(V_1\right)\right] \\
& - c_2h^2  b^{-2} \be_3^\top \mathbf{D}_{ b, v_1^0, 3}^{-1} K_{b, v_1^0}\left(V_1\right) \bg_{b, v_1^0, 3}\left(V_1\right) \bg_{b, v_1^0, 3}^\top\left(V_1\right) \mathbf{D}_{ b, v_1^0, 3}^{-1} \mathbb{E}\left[\bg_{b, v_1^0, 3}\left(V_1\right) K_{b, v_1^0}\left(V_1\right)  \tau_1\left(V_1\right)\right]
\end{aligned}
\]
\end{theorem}
Let $\hat{\phi}_{ h, b, v_1^0}^{\text{cate}}$ be the empirical influence function and define $\hat{\sigma}_{h, b}^2\left(v_1^0\right):=h \MP_n\left[(\hat{\phi}_{ h, b, v_1^0}^{\text{cate}})^2\right]$ as the variance estimator. Let $\mathcal{V}_1$ denote the support of $V_1$ (or the range of $V_1$ that we want to construct confidence band on) and $\mathcal{V}_n$ denote a suitable grid approximation of $\mathcal{V}_1$. Then under regularity conditions similar to \cite{takatsu2022debiased}, one can show
\[
\sup _{t \in \mathbb{R}}\left|\MP\left(\sup _{v_1^0 \in \mathcal{V}_1}(n h)^{1 / 2}\left|\frac{\hat{\tau}_{1}^{\text{de}}\left(v_1^0\right)-\tau_1\left(v_1^0\right)}{\hat{\sigma}_{h, b}\left(v_1^0\right)}\right| \leq t\right)-\MP\left(\max _{v_1^0 \in \mathcal{V}_n}\left|Z_{n}\left(v_1^0\right)\right| \leq t \big | \bZ_1 ,\dots, \bZ_n \right)\right|=o_p(1),
\]
where $Z_{n}$ is a multivariate Gaussian vector on $\mathcal{V}_n$ with covariance given by
\[
\operatorname{Cov}\left(Z_{n}\left(x\right), Z_{n}\left(y\right)\right)=h \mathbb{P}_n\left(\hat{\phi}_{ h, b, x}^{\text{cate}}\hat{\phi}_{ h, b, y}^{\text{cate}}\right) /\left[\hat{\sigma}_{ h, b}\left(x\right) \hat{\sigma}_{ h, b}\left(y\right)\right].
\]
The proof of this statement is very similar to \cite{takatsu2022debiased} and is omitted here. Hence one can use the quantile of $\max _{v_1^0 \in \mathcal{V}_n}\left|Z_{n}\left(v_1^0\right)\right|$ (which can be obtained by simulations) to approximate the quantile of
\[
\sup _{v_1^0 \in \mathcal{V}_1}(n h)^{1 / 2}\left|\frac{\hat{\tau}_{1}^{\text{de}}\left(v_1^0\right)-\tau_1\left(v_1^0\right)}{\hat{\sigma}_{h, b}\left(v_1^0\right)}\right|
\]
and construct asymptotic valid confidence band of $\tau_1(v_1)$ based on it. In the following section, we prove Theorem \ref{thm:EIF-cate}.
\subsection{Proof of Theorem \ref{thm:EIF-cate}}

We will only consider the EIF of the functional
\[
\tau_h(v_1^0;\MP) = \be_1^{\top} \bD_{h, v_1^0, 1}^{-1} \ME_{\MP}[\bg_{h, v_1^0, 1} (V_1) K_{h, v_1^0}(V_1) \tau_1(V_1)]
\]
as the construction of the EIF for the other term is entirely analogous. We will assume $V_1$ is the first coordinate in $\bX$. Let $p_{\epsilon}(\bo) = p(\bo ; \epsilon)$ denote a parametric submodel with parameter $\epsilon \in \mathbb{R}$. Denote $S_{\epsilon} (\bU | \bW) = \left.\{\partial \log p_{\epsilon}(\bU|\bW) / \partial \varepsilon\}\right|_{\varepsilon=0}$ as the scores on the parametric submodels $p_{\epsilon}(\bu|\mathbf{w})$, where $\bU$ and $\bW$ are some set of random variables. By definition, (note that $\bD_{h, v_1^0, 1}$ and $\tau_1(V_1)$ also depend on $\MP$)
\[
\begin{aligned}
    \frac{\partial \tau_h(v_1^0;\MP_{\epsilon})}{\partial \epsilon} \bigg|_{\epsilon=0} =&\, \be_1^{\top} \frac{\partial \bD_{h,v_1^0, 1, \epsilon}^{-1}}{\partial \epsilon} \bigg|_{\epsilon=0} \ME[\bg_{h, v_1^0, 1} (V_1) K_{h, v_1^0}(V_1) \tau_1(V_1)] \\
    &\, + \be_1^{\top} \bD_{h, v_1^0, 1}^{-1}  \frac{\partial \ME_{\MP_{\epsilon}}[\bg_{h, v_1^0, 1} (V_1) K_{h, v_1^0}(V_1) \tau_1(V_1)] }{\partial \epsilon} \bigg|_{\epsilon=0} \\
    & \, + \be_1^{\top} \bD_{h, v_1^0, 1}^{-1} \ME \left[ \bg_{h, v_1^0, 1} (V_1) K_{h, v_1^0}(V_1) \frac{\partial \tau_{1,\epsilon}(V_1)}{\partial \epsilon} \bigg|_{\epsilon=0}\right]\\
    := &\, I_1 + I_2 + I_3.
\end{aligned}
\]
where $\tau_{1,\epsilon}(V_1)$ is the value of $\tau_{1}(V_1)$ evaluated at distribution $\MP_\epsilon$. We have
\[
\begin{aligned}
I_1 = &\,  -\be_1^{\top} \bD_{h, v_1^0, 1}^{-1} \frac{\partial \bD_{h,v_1^0, 1, \epsilon}}{\partial \epsilon}\bigg|_{\epsilon=0} \bD_{h, v_1^0, 1}^{-1} \ME[\bg_{h, v_1^0, 1} (V_1) K_{h, v_1^0}(V_1) \tau_1(V_1)]\\
= &\, -\be_1^{\top} \bD_{h, v_1^0, 1}^{-1} \ME[\bg_{h, v_1^0, 1} (V_1) K_{h, v_1^0}(V_1)\bg_{h, v_1^0, 1}^{\top} (V_1) S_{\epsilon}(V_1)] \bD_{h, v_1^0, 1}^{-1} \ME[\bg_{h, v_1^0, 1} (V_1) K_{h, v_1^0}(V_1) \tau_1(V_1)].
\end{aligned}
\]
For term $I_2$ we have
\[
I_2 = \be_1^{\top}\bD_{h, v_1^0, 1}^{-1} \ME[\bg_{h, v_1^0, 1} (V_1) K_{h, v_1^0}(V_1) \tau_1(V_1)S_{\epsilon}(V_1)].
\]
For term $I_3$, we first consider (let $\bx_{-1}$ be the covariates excluding $v_1$)
\[
\begin{aligned}
  &\, \ME \left[ \bg_{h, v_1^0, 1} (V_1) K_{h, v_1^0}(V_1) \frac{\partial \ME_{\epsilon}[\mu_{a,\epsilon}(\bX)|V_1]}{\partial \epsilon} \bigg|_{\epsilon=0}\right] \\
  = &\, \int \bg_{h, v_1^0, 1} (v_1) K_{h, v_1^0}(v_1)  \iint y S_{\epsilon}(y|A=a,\bx) p(y|A=a,\bx)dy \, p(\bx_{-1}|v_1) d \bx_{-1} \, p(v_1) dv_1 \\
  &\, + \int \bg_{h, v_1^0, 1} (v_1) K_{h, v_1^0}(v_1)  \iint y p(y|A=a,\bx)dy \, p(\bx_{-1}|v_1) S_{\epsilon}(\bx_{-1}|v_1) d \bx_{-1} \, p(v_1) dv_1.
\end{aligned}
\]
For the first term we have ($\pi_a (\bX) = \MP(A=a|\bX)$)
\[
\begin{aligned}
    &\, \int \bg_{h, v_1^0, 1} (v_1) K_{h, v_1^0}(v_1)  \iint y S_{\epsilon}(y|A=a,\bx) p(y|A=a,\bx)dy \, p(\bx_{-1}|v_1) d \bx_{-1} \, p(v_1) dv_1\\
    = &\, \ME \left\{\bg_{h, v_1^0, 1} (V_1) K_{h, v_1^0}(V_1)\ME [\ME[Y S_{\epsilon}(Y|A=a,\bX) |A=a, \bX]|V_1]  \right\} \\
    =& \,\ME \left\{\bg_{h, v_1^0, 1} (V_1) K_{h, v_1^0}(V_1)\ME [\ME[(Y-\mu_a(\bX) )S_{\epsilon}(Y|A=a,\bX)|A=a, \bX]|V_1]  \right\} \\
     =& \,\ME \left\{\bg_{h, v_1^0, 1} (V_1) K_{h, v_1^0}(V_1)\ME \left[ \frac{I(A=a)}{\pi_a(\bX)}\ME[(Y-\mu_a(\bX) )S_{\epsilon}(Y|A,\bX)|A, \bX]|V_1 \right]  \right\}\\
    = & \, \ME \left[ \bg_{h, v_1^0, 1} (V_1) K_{h, v_1^0}(V_1)\frac{I(A=a)}{\pi_a(\bX)} (Y-\mu_a(\bX) )S_{\epsilon}(Y|A,\bX) \right] \\
    = &\, \ME \left[ \bg_{h, v_1^0, 1} (V_1) K_{h, v_1^0}(V_1)\frac{I(A=a)}{\pi_a(\bX)} (Y-\mu_a(\bX) )S_{\epsilon}(Y,A,\bX) \right]
\end{aligned}
\]
where in the second equation we use $\ME[S(Y|A=a, \bX)|A=a,\bX] = 0$. The third and fourth equation follow from property of conditional expectation and the last equation follows from the fact $\ME \left[ \bg_{h, v_1^0, 1} (V_1) K_{h, v_1^0}(V_1)\frac{I(A=a)}{\pi_a(\bX)} (Y-\mu_a(\bX) )S_{\epsilon}(A,\bX) \right] = 0$ (recall $S_{\epsilon}(Y,A,\bX) = S_{\epsilon}(A,\bX) + S_{\epsilon}(Y|A,\bX)$).
For the second term we have (define $\eta_a(V_1) = \ME[\mu_a(\bX)|V_1]$)
\[
\begin{aligned}
&\,\int \bg_{h, v_1^0, 1} (v_1) K_{h, v_1^0}(v_1)  \iint y p(y|A=a,\bx)dy \, p(\bx_{-1}|v_1) S_{\epsilon}(\bx_{-1}|v_1) d \bx_{-1} \, p(v_1) dv_1 \\
= &\, \ME \left \{ \bg_{h, v_1^0, 1} (V_1) K_{h, v_1^0}(V_1) \ME[\mu_a(\bX)S_{\epsilon}(\bX_{-1}|V_1)|V_1] \right \} \\
= &\, \ME \left \{ \bg_{h, v_1^0, 1} (V_1) K_{h, v_1^0}(V_1) \ME[(\mu_a(\bX)-\eta_a(V_1))S_{\epsilon}(\bX_{-1}|V_1)|V_1] \right \} \\
 = &\, \ME[\bg_{h, v_1^0, 1} (V_1) K_{h, v_1^0}(V_1) (\mu_a(\bX)-\eta_a(V_1))S_{\epsilon}(\bX_{-1}|V_1)] \\
 = &\, \ME[\bg_{h, v_1^0, 1} (V_1) K_{h, v_1^0}(V_1) (\mu_a(\bX)-\eta_a(V_1))S_{\epsilon}(\bX)]
\end{aligned}
\]
where the second equation follows from $\ME[S_{\epsilon}(\bX_{-1}|V_1)|V_1]=0$, the third equation follows from property of conditional expectation and the last equation follows from $\ME[\bg_{h, v_1^0, 1} (V_1) K_{h, v_1^0}(V_1) (\mu_a(\bX)-\eta_a(V_1))S_{\epsilon}(V_1)]=0$. Combining the results for $A=1$ and $A=0$, we have
\[
\begin{aligned}
    I_3 = &\,\be_1^{\top} \bD_{h, v_1^0, 1}^{-1} \ME \left[ \bg_{h, v_1^0, 1} (V_1) K_{h, v_1^0}(V_1) \frac{\partial \tau_{1,\epsilon}(V_1)}{\partial \epsilon} \bigg|_{\epsilon=0}\right] \\
    = &\, \be_1^{\top} \bD_{h, v_1^0, 1}^{-1} \ME\left[ \bg_{h, v_1^0, 1} (V_1) K_{h, v_1^0}(V_1) ({\varphi}^{\text{cate}}(\bZ) - \tau_1(V_1)) S_{\epsilon}(Y,A,\bX)\right].
\end{aligned}
\]
Add $I_1, I_2, I_3$ together, we have
\[
\frac{\partial \tau_h(v_1^0;\MP_{\epsilon})}{\partial \epsilon} \bigg|_{\epsilon=0} = \be_1^{\top} \bD_{h, v_1^0, 1}^{-1} \ME[\bg_{h, v_1^0, 1} (V_1) K_{h, v_1^0}(V_1) ({\varphi}^{\text{cate}}(\bZ) - \gamma_{h,v_1^0}^{\text{cate}}(V_1)) S_{\epsilon}(Y,A,\bX)],
\]
where 
\[
\gamma_{h,v_1^0}^{\text{cate}}(V_1) = \bg_{h, v_1^0, 1} (V_1)^{\top} \bD_{h, v_1^0, 1}^{-1} \ME[\bg_{h, v_1^0, 1} (V_1) K_{h, v_1^0}(V_1) \tau_1(V_1)].
\]
Let 
\[
\phi_{h,v_1^0}^{\text{cate}}(\bZ) = \be_1^{\top} \bD_{h, v_1^0, 1}^{-1} \bg_{h, v_1^0, 1} (V_1) K_{h, v_1^0}(V_1) ({\varphi}^{\text{cate}}(\bZ) - \gamma_{h,v_1^0}^{\text{cate}}(V_1)).
\]
Note that
\[
\ME[\be_1^{\top} \bD_{h, v_1^0, 1}^{-1} \bg_{h, v_1^0, 1} (V_1) K_{h, v_1^0}(V_1)\gamma_{h,v_1^0}^{\text{cate}}(V_1)] = \be_1^{\top} \bD_{h, v_1^0, 1}^{-1} \ME[\bg_{h, v_1^0, 1} (V_1) K_{h, v_1^0}(V_1) \tau_1(V_1)]
\]
and
\[
\ME[\be_1^{\top} \bD_{h, v_1^0, 1}^{-1} \bg_{h, v_1^0, 1} (V_1) K_{h, v_1^0}(V_1) {\varphi}^{\text{cate}}(\bZ)] = \be_1^{\top} \bD_{h, v_1^0, 1}^{-1} \ME[\bg_{h, v_1^0, 1} (V_1) K_{h, v_1^0}(V_1) \tau_1(V_1)],
\]
so $\phi_{h,v_1^0}^{\text{cate}}(\bZ)$ is centered and is the efficient influence function of $\tau_h(v_1^0;\MP)$. The efficient influence function of $\be_3^{\top} \bD_{b, v_1^0, 3}^{-1} \ME[\bg_{b, v_1^0, 3} (V_1) K_{b, v_1^0}(V_1) \tau_1(V_1)]$ can be similarly obtained and the result of the theorem follows.

\section{Results for the partial dependence function \label{appendix:partial_dependence}}
\begin{theorem}\label{thm:pd_dr}
Let $\bar{\eta} = \left(\bar{\pi}, \bar{\mu}_0, \bar{\mu}_1, \bar{f}_{j|-j}\right)$ be nuisance functions that may not necessarily equal the real ones $\eta$. We have:
\[
\ME[\varphi_{j}^{\text{pd}}\left(\bZ; \bar{\eta} \right) | V_j = v_j] = \theta_j(v_j)
\]
if either $(\bar{\mu}_0, \bar{\mu}_1) = (\mu_0, \mu_1)$ or $\left(\bar{\pi}, \bar{f}_{j|-j}\right) = \left(\pi, f_{j|-j}\right)$ holds.
\end{theorem}
Theorem \ref{thm:pd_dr} shows that the pseudo-oucome used to construct the DR-Learner estimator of the partial dependence function is doubly-robust. In fact, $\varphi_{j}^{\text{pd}}\left(\bZ; \bar{\eta} \right)$ has the correct conditional expectation if we correctly specify either the outcome regression model $(\mu_0, \mu_1)$ or the propensity score and conditional density $\left(\pi, f_{j|-j}\right)$. This motivates us to apply Algorithm \ref{alg:cate-estimation} to estimate $\theta_j(v_j)$ with suitable modifications. In the first step we model the nuisance functions $\eta$ that appear in the pseudo-outcome $\varphi_{j}^{\text{pd}}(\bZ;\eta)$ with flexible non-parametric or machine learning methods. For this parameter, one needs to further regress $\tau_x(\bX) $ on $\bV$ to estimate $\tau_v(\bV) = \ME[\tau_x(\bX) \mid \bV]$ and model the conditional density $f_{j|-j}(v_j|\bv_{-j})$. Once we have these estimates, the marginal density $f_j(v_j)$ can be estimated by the following estimator:
\[
\hat{f}_j(v_j) =  \int_{\mathcal{V}_{-j}} \hat{f}_{j|-j}(v_j|\bv_{-j}) d \MP_n (\bv_{-j}) = \frac{1}{n}\sum_{i=1}^n \hat{f}_{j|-j}(v_j|\bv_{i,-j}).
\]
Similarly to estimate $\theta_j(v_j)$ in constructing the pesudo-outcome, we can use the estimator for $\tau_v(\bv)$ as
\[
\hat{\theta}_j(v_j) = \int_{\mathcal{V}_{-j}} \hat{\tau}_v(v_j, \bv_{-j}) d \MP_n (\bv_{-j}) = \frac{1}{n} \sum_{i=1}^n \hat{\tau}_v(v_j,\bv_{i,-j}).
\]
One can proceed similarly as in Appendix \ref{appendix:individual-cate-band} to construct uniform confidence band by using the influence function of the localized functional $\theta_{h, b,j}\left(v_j^0\right)$, summarized in the following theorem.

\begin{theorem}\label{thm:EIF-pd}
For each $h, b > 0$ and $v_1^0 \in \mathcal{V}_1$, consider a nonparametric model class consisting of $\MP$ such that under $\MP$, $\ME [Y^2] < \infty$ and $f_{1|-1}(v_1 | \bv_{-1}) > 0$ for all $v_1 \in [v_1^0 - \min(h,b), v_1^0 + \min(h,b)],\bv_{-1} \in \mathcal{V}_{-1}$, the efficient influence function of $\theta_{h, b,1}\left(v_1^0\right)=\mathbb{E}\left[\Gamma_{ h, b, v_1^0}\left(V_1\right) \theta_1(V_1)\right]$ under this model class is
    \[
    \begin{aligned}
& \phi_{h, b, v_1^0}^{\text{pd}}(\bZ)=\Gamma_{h, b, v_1^0}\left(V_1\right) \varphi_{1}^{\text{pd}}(\bZ) -\gamma_{h, b, v_1^0}^{\text{pd}}\left(V_1\right)+\int \Gamma_{ h, b, v_1^0}\left(\bar{v}_1\right)\left[\tau_v\left(\bar{v}_1, \mathbf{V}_{-1}\right)-\theta\left(\bar{v}_1\right)\right] d \MP\left(\bar{v}_1\right), \text { where } \\
& \gamma_{h, b, v_1^0}^{\text{pd}}\left(V_1\right)=\be_1^\top \mathbf{D}_{h, v_1^0, 1}^{-1} K_{h, v_1^0}\left(V_1\right) \bg_{h, v_1^0, 1}\left(V_1\right) \bg_{h, v_1^0, 1}^{\top}\left(V_1\right) \mathbf{D}_{h, v_1^0, 1}^{-1} \mathbb{E}\left[ \bg_{h, v_1^0, 1}\left(V_1\right) K_{h, v_1^0}\left(V_1\right) \theta_1\left(V_1\right)\right] \\
& - c_2 h^2b^{-2} \be_3^\top  \mathbf{D}_{b, v_1^0, 3}^{-1} K_{b, v_1^0}\left(V_1\right) \bg_{b, v_1^0, 3}\left(V_1\right) \bg_{b, v_1^0, 3}^\top\left(V_1\right) \mathbf{D}_{ b, v_1^0, 3}^{-1} \mathbb{E}\left[\bg_{b, v_1^0, 3}\left(V_1\right)  K_{b, v_1^0}\left(V_1\right) \theta_1\left(V_1\right)\right] \\
&
\end{aligned}
    \]
\end{theorem}
In the next two sections, we prove Theorems \ref{thm:pd_dr} and \ref{thm:EIF-pd}.
\subsection{Proof of Theorem \ref{thm:pd_dr}}

We first show
\[
\ME \left[ \frac{(A-\bar{\pi}(\bX))(Y-\bar{\mu}_A(\bX))}{\bar{\pi}(\bX)(1-\bar{\pi}(\bX))} + \bar{\tau}_x(\bX) \bigg | \bV \right] = \tau_v(\bV)
\]
under the assumption of either $(\bar{\mu}_0, \bar{\mu}_1) = ({\mu}_0, {\mu}_1)$ or $\bar{\pi} = \pi$. By direct calculations we have
\begin{equation}\label{eq:pseudo}
    \frac{(A-\bar{\pi}(\bX))(Y-\bar{\mu}_A(\bX))}{\bar{\pi}(\bX)(1-\bar{\pi}(\bX))} + \bar{\tau}_x(\bX) = \frac{A(Y-\bar{\mu}_1(\bX))}{\bar{\pi}(\bX)} + \bar{\mu}_1(\bX) - \left[ \frac{(1-A)(Y-\bar{\mu}_0(\bX))}{1-\bar{\pi}(\bX)} + \bar{\mu}_0(\bX)  \right].
\end{equation}
For each treatment $A=a$, we have 
\[
\begin{aligned}
   & \, \ME \left[ \frac{I(A=a)(Y-\bar{\mu}_a(\bX))}{\bar{\pi}_a(\bX)} + \bar{\mu}_a(\bX) - \mu_a(\bX) \bigg |\bV \right]  \\
   = &\, \ME \left[ (\bar{\mu}_a(\bX) - \mu_a(\bX)) \left( 1-\frac{\pi_a(\bX)}{\bar{\pi}_a (\bX)}  \right) \bigg | \bV \right] .
\end{aligned}
\]
So either $\bar{\mu}_a = \mu_a$ or $\bar{\pi}_a = \pi_a$ yields 
\[
\ME \left[ \frac{I(A=a)(Y-\bar{\mu}_a(\bX))}{\bar{\pi}_a(\bX)} + \bar{\mu}_a(\bX)  \bigg |\bV \right] =  \ME[\mu_a(\bX) | \bV].
\]
This together with \eqref{eq:pseudo} shows either $(\bar{\mu}_0, \bar{\mu}_1) = ({\mu}_0, {\mu}_1)$ or $\bar{\pi} = \pi$ implies
\[
\ME \left[ \frac{(A-\bar{\pi}(\bX))(Y-\bar{\mu}_A(\bX))}{\bar{\pi}(\bX)(1-\bar{\pi}(\bX))} + \bar{\tau}_x(\bX) \bigg | \bV \right] = \tau_v(\bV).
\]
Hence under either $(\bar{\mu}_0, \bar{\mu}_1) = (\mu_0, \mu_1)$ or $\left(\bar{\pi}, \bar{f}_{j|-j}\right) = \left(\pi, f_{j|-j}\right)$, we have
\[
\ME[{\varphi}_j^{\text{pd}}\left(\bZ; \bar{\eta} \right) | \bV] = \left( \ME[\tau_x(\bX ) | \bV] - \ME[\bar{\tau}_x(\bX)|\bV] \right)\frac{\int \bar{f}_{j|-j}(V_j | \bv_{-j}) d \MP(\bv_{-j})}{\bar{f}_{j|-j}(V_j | \bV_{-j})} + \int \bar{\tau}_v(V_j, \bv_{-j}) d \MP(\bv_{-j}),
\]
where $\bar{\tau}_v(\bV) = \ME[\bar{\tau}_x(\bX) | \bV]$. Now we consider two cases separately. First assume $(\bar{\mu}_0, \bar{\mu}_1) = (\mu_0, \mu_1)$, we have
\[
\ME[{\varphi}_j^{\text{pd}}\left(\bZ; \bar{\eta} \right) | \bV] = \int {\tau}_v(V_j, \bv_{-j}) d \MP(\bv_{-j}) = \theta_j(V_j),
\]
which implies the claim of Theorem \ref{thm:pd_dr}. Under the other assumption $\left(\bar{\pi}, \bar{f}_{j|-j}\right) = \left(\pi, f_{j|-j}\right)$, we have
\[
\ME[{\varphi}_j^{\text{pd}}\left(\bZ; \bar{\eta} \right) | \bV] = \left( \tau_v(\bV) - \bar{\tau}_v(\bV) \right)\frac{f_j(V_j)}{{f}_{j|-j}(V_j | \bV_{-j})} + \int \bar{\tau}_v(V_j, \bv_{-j}) d \MP(\bv_{-j}), 
\]
where $f_j(V_j)=\int {f}_{j|-j}(V_j | \bv_{-j}) d \MP(\bv_{-j})$. Hence we conclude
\[
\begin{aligned}
\ME[{\varphi}_j^{\text{pd}}\left(\bZ; \bar{\eta} \right) | V_j= v_j] = &\, \int (\tau_v(\bv) - \bar{\tau}_v(\bv)) \frac{f_j(v_j)}{{f}_{j|-j}(v_j | \bv_{-j})} d \MP(\bv_{-j}|v_j) + \int \bar{\tau}_v(v_j, \bv_{-j}) d \MP(\bv_{-j}) \\
= & \, \int (\tau_v(\bv) - \bar{\tau}_v(\bv)) \frac{f_j(v_j)}{{f}_{j|-j}(v_j | \bv_{-j})} \frac{f_{j|-j}(v_j|\bv_{-j})}{f_j(v_j)} d \MP(\bv_{-j}) + \int \bar{\tau}_v(\bv) d \MP(\bv_{-j}) \\
= & \, \int (\tau_v(\bv) - \bar{\tau}_v(\bv))  d \MP(\bv_{-j}) + \int \bar{\tau}_v(\bv) d \MP(\bv_{-j}) \\
= & \, \theta_j(v_j),
\end{aligned}
\]
where in the second equality we use the fact 
\[
d \MP(\bv_{-j}|v_j) = \frac{f_{j|-j}(v_j|\bv_{-j})}{f_j(v_j)} d \MP(\bv_{-j}).
\]

\subsection{Proof of Theorem \ref{thm:EIF-pd}}

We will use the same notation as in proof of Theorem \ref{thm:EIF-cate}. Consider the EIF of functional
\[
\theta_h(v_1^0;\MP) = \be_1^{\top} \bD_{h, v_1^0, 1}^{-1} \ME_{\MP}[\bg_{h, v_1^0, 1} (V_1) K_{h, v_1^0}(V_1) \theta_1(V_1)]
\]
By definition (note that $\bD_{h, v_1^0, 1}$ and $\theta_1(V_1)$ also depend on $\MP$)
\[
\begin{aligned}
    \frac{\partial \theta_h(v_1^0;\MP_{\epsilon})}{\partial \epsilon} \bigg|_{\epsilon=0} =&\, \be_1^{\top} \frac{\partial \bD_{h,v_1^0, 1, \epsilon}^{-1}}{\partial \epsilon} \bigg|_{\epsilon=0} \ME[\bg_{h, v_1^0, 1} (V_1) K_{h, v_1^0}(V_1) \theta_1(V_1)] \\
    &\, + \be_1^{\top} \bD_{h, v_1^0, 1}^{-1}  \frac{\partial \ME_{\MP_{\epsilon}}[\bg_{h, v_1^0, 1} (V_1) K_{h, v_1^0}(V_1) \theta_1(V_1)] }{\partial \epsilon} \bigg|_{\epsilon=0} \\
    & \, + \be_1^{\top} \bD_{h, v_1^0, 1}^{-1} \ME \left[ \bg_{h, v_1^0, 1} (V_1) K_{h, v_1^0}(V_1) \frac{\partial \theta_{1,\epsilon}(V_1)}{\partial \epsilon} \bigg|_{\epsilon=0}\right]\\
    := &\, I_1 + I_2 + I_3,
\end{aligned}
\]
where $\theta_{1,\epsilon}(V_1)$ is the value of $\theta_{1}(V_1)$ evaluated at distribution $\MP_\epsilon$.
We have
\[
\begin{aligned}
I_1 = &\,  -\be_1^{\top} \bD_{h, v_1^0, 1}^{-1} \frac{\partial \bD_{h,v_1^0, 1, \epsilon}}{\partial \epsilon} \bigg|_{\epsilon=0} \bD_{h, v_1^0, 1}^{-1} \ME[\bg_{h, v_1^0, 1} (V_1) K_{h, v_1^0}(V_1) \theta_1(V_1)]\\
= &\, -\be_1^{\top} \bD_{h, v_1^0, 1}^{-1} \ME[\bg_{h, v_1^0, 1} (V_1) K_{h, v_1^0}(V_1)\bg_{h, v_1^0, 1}^{\top} (V_1) S_{\epsilon}(V_1)] \bD_{h, v_1^0, 1}^{-1} \ME[\bg_{h, v_1^0, 1} (V_1) K_{h, v_1^0}(V_1) \theta_1(V_1)].
\end{aligned}
\]
For term $I_2$ we have
\[
I_2 = \be_1^{\top}\bD_{h, v_1^0, 1}^{-1} \ME[\bg_{h, v_1^0, 1} (V_1) K_{h, v_1^0}(V_1) \theta_1(V_1)S_{\epsilon}(V_1)].
\]
For term $I_3$, we first consider (let $\bX = (\bW, \bV)$)
\[
\begin{aligned}
  &\, \ME \left[ \bg_{h, v_1^0, 1} (V_1) K_{h, v_1^0}(V_1) \frac{\partial  \int \ME_{\epsilon}[\mu_{a,\epsilon}(\bX)|\bV] f_{-1,\epsilon}(\bV_{-1}) d \bV_{-1}}  {\partial \epsilon} \bigg|_{\epsilon=0}\right] \\
  = &\, \int \bg_{h, v_1^0, 1} (v_1) K_{h, v_1^0}(v_1)  \iiint y S_{\epsilon}(y|A=a,\bx) p(y|A=a,\bx)dy \, 
  p(\bw|\bv) d \bw\, f_{-1}(\bv_{-1}) d \bv_{-1} \, f_1(v_1) dv_1 \\
  &\, + \int \bg_{h, v_1^0, 1} (v_1) K_{h, v_1^0}(v_1)  \iiint y p(y|A=a,\bx)dy \, S_{\epsilon}(\bw|\bv) p(\bw|\bv) d \bw \, f_{-1}(\bv_{-1}) d\bv_{-1} \, f_1(v_1) dv_1 \\
  &\, +  \int \bg_{h, v_1^0, 1} (v_1) K_{h, v_1^0}(v_1)  \iiint y p(y|A=a,\bx)dy \,  p(\bw|\bv) d \bw \, S_{\epsilon}(\bv_{-1}) f_{-1}(\bv_{-1}) d\bv_{-1} \, f_1(v_1) dv_1
\end{aligned}
\]
For the first term we have
\[
\begin{aligned}
    &\, \int \bg_{h, v_1^0, 1} (v_1) K_{h, v_1^0}(v_1)  \iiint y S_{\epsilon}(y|A=a,\bx) p(y|A=a,\bx)dy \, 
  p(\bw|\bv) d \bw\, f_{-1}(\bv_{-1}) d \bv_{-1} \, f_1(v_1) dv_1 \\
    = &\, \ME \left\{\bg_{h, v_1^0, 1} (V_1) K_{h, v_1^0}(V_1) \frac{f_1(V_1)}{f_{1|-1}(V_1|\bV_{-1})}\ME [\ME[Y S_{\epsilon}(Y|A=a,\bX) |A=a, \bX]|\bV]  \right\} \\
    =& \,\ME \left\{\bg_{h, v_1^0, 1} (V_1) K_{h, v_1^0}(V_1) \frac{f_1(V_1)}{f_{1|-1}(V_1|\bV_{-1})} \ME [\ME[(Y-\mu_a(\bX) )S_{\epsilon}(Y|A=a,\bX)|A=a, \bX]|V_1]  \right\} \\
     =& \,\ME \left\{\bg_{h, v_1^0, 1} (V_1) K_{h, v_1^0}(V_1) \frac{f_1(V_1)}{f_{1|-1}(V_1|\bV_{-1})} \ME \left[ \frac{I(A=a)}{\pi_a(\bX)}\ME[(Y-\mu_a(\bX) )S_{\epsilon}(Y|A,\bX)|A, \bX]|V_1 \right]  \right\}\\
    = & \, \ME \left[ \bg_{h, v_1^0, 1} (V_1) K_{h, v_1^0}(V_1)\frac{I(A=a)}{\pi_a(\bX)} (Y-\mu_a(\bX) ) \frac{f_1(V_1)}{f_{1|-1}(V_1|\bV_{-1})}S_{\epsilon}(Y|A,\bX) \right] \\
    = &\, \ME \left[ \bg_{h, v_1^0, 1} (V_1) K_{h, v_1^0}(V_1)\frac{I(A=a)}{\pi_a(\bX)} (Y-\mu_a(\bX) )S_{\epsilon}(Y,A,\bX) \right]
\end{aligned}
\]
where in the first equation we note $f_{-1}(\bv_{-1}) f_1(v_1) = \frac{f_1(v_1) f(v_1, \bv_{-1})}{f_{1|-1}(v_1|\bv_{-1})}$, in the second equation we use $\ME[S(Y|A=a, \bX)|A=a,\bX] = 0$. The third and fourth equation follow from property of conditional expectation and the last equation follows from the fact 
$$\ME \left[ \bg_{h, v_1^0, 1} (V_1) K_{h, v_1^0}(V_1)\frac{I(A=a)}{\pi_a(\bX)} (Y-\mu_a(\bX) )S_{\epsilon}(A,\bX) \right]=0$$
We used the fact that $S_{\epsilon}(Y,A,\bX) = S_{\epsilon}(A,\bX) + S_{\epsilon}(Y|A,\bX)$.
For the second term we have (define $\beta_a(\bV) = \ME[\mu_a(\bX)|\bV]$)
\[
\begin{aligned}
&\,\int \bg_{h, v_1^0, 1} (v_1) K_{h, v_1^0}(v_1)  \iiint y p(y|A=a,\bx)dy \, S_{\epsilon}(\bw|\bv) p(\bw|\bv) d \bw \, f_{-1}(\bv_{-1}) d\bv_{-1} \, f_1(v_1) dv_1 \\
= &\, \ME \left \{ \bg_{h, v_1^0, 1} (V_1) K_{h, v_1^0}(V_1) \ME[\mu_a(\bX)S_{\epsilon}(\bW|\bV)|\bV] \frac{f_1(V_1)}{f_{1|-1}(V_1|\bV_{-1})} \right \} \\
= &\, \ME \left \{ \bg_{h, v_1^0, 1} (V_1) K_{h, v_1^0}(V_1) \ME[(\mu_a(\bX)-\beta_a(\bV))S_{\epsilon}(\bW|\bV)|\bV] \frac{f_1(V_1)}{f_{1|-1}(V_1|\bV_{-1})}  \right \} \\
 = &\, \ME \left[\bg_{h, v_1^0, 1} (V_1) K_{h, v_1^0}(V_1) (\mu_a(\bX)-\beta_a(\bV))S_{\epsilon}(\bW|\bV) \frac{f_1(V_1)}{f_{1|-1}(V_1|\bV_{-1})} \right] \\
 = &\, \ME \left[\bg_{h, v_1^0, 1} (V_1) K_{h, v_1^0}(V_1) (\mu_a(\bX)-\beta_a(\bV)) \frac{f_1(V_1)}{f_{1|-1}(V_1|\bV_{-1})} S_{\epsilon}(\bX) \right]
\end{aligned}
\]
where the second equation follows from $\ME[S_{\epsilon}(\bW|\bV)|\bV]=0$, the third equation follows from property of conditional expectation and the last equation follows from
\[
\ME \left[\bg_{h, v_1^0, 1} (V_1) K_{h, v_1^0}(V_1) (\mu_a(\bX)-\beta_a(\bV)) \frac{f_1(V_1)}{f_{1|-1}(V_1|\bV_{-1})} S_{\epsilon}(\bV) \right] = 0.
\]
For the third term, we have
\[
\begin{aligned}
    &\, \int \bg_{h, v_1^0, 1} (v_1) K_{h, v_1^0}(v_1)  \iiint y p(y|A=a,\bx)dy \,  p(\bw|\bv) d \bw \, S_{\epsilon}(\bv_{-1}) f_{-1}(\bv_{-1}) d\bv_{-1} \, f_1(v_1) dv_1 \\
    =&\, \ME \left [ \int \bg_{h, v_1^0, 1}(v_1) K_{h,v_1^0}(v_1) \beta_a(v_1, \bV_{-1}) f_1(v_1) d v_1 \, S_{\epsilon}(\bV_{-1})  \right] \\
    = &\, \ME \left [ \int \bg_{h, v_1^0, 1}(v_1) K_{h,v_1^0}(v_1) \beta_a(v_1, \bV_{-1}) f_1(v_1) d v_1 \, S_{\epsilon}(Y,A,\bX)  \right]
\end{aligned}
\]
For notational simplicity, define 
\[
\br(\bv_{-1}) = \int \bg_{h, v_1^0, 1}(v_1) K_{h, {v_1^0}} (v_1) \tau_v(v_1, \bv_{-1}) f_1(v_1) d v_1 ,
\]
Combining the results for $A=1$ and $A=0$, we have
\[
\begin{aligned}
    I_3 = &\,\be_1^{\top} \bD_{h, v_1^0, 1}^{-1} \ME \left[ \bg_{h, v_1^0, 1} (V_1) K_{h, v_1^0}(V_1) \frac{\partial \theta_{1,\epsilon}(V_1)}{\partial \epsilon} \bigg|_{\epsilon=0}\right] \\
    = &\, \be_1^{\top} \bD_{h, v_1^0, 1}^{-1} \ME\left\{ \left[\bg_{h, v_1^0, 1} (V_1) K_{h, v_1^0}(V_1) \left( \frac{(A-\pi(\bX))(Y-\mu_A(\bX))}{\pi(\bX)(1-\pi(\bX))} + \tau_x(\bX) - \tau_v(\bV) \right)   \right. \right.   \\
     & \, \left. \left. \frac{f_1(V_1)}{f_{1|-1}(V_1|\bV_{-1})}  + \br(\bV_{-1}) \right] 
  S_{\epsilon}(Y,A,\bX) \right\}. 
\end{aligned}
\]
Adding $I_1, I_2, I_3$ together, we have
\[
\frac{\partial \theta_h(v_1^0;\MP_{\epsilon})}{\partial \epsilon} \bigg|_{\epsilon=0} = \be_1^{\top} \bD_{h, v_1^0, 1}^{-1} \ME\left\{ [\bg_{h, v_1^0, 1} (V_1) K_{h, v_1^0}(V_1) ({\varphi}_1^{\text{pd}}(\bZ) - \gamma_{h,v_1^0}^{\text{pd}}(V_1)) + \br(\bV_{-1}) ]S_{\epsilon}(Y,A,\bX) \right\},
\]
where 
\[
\gamma_{h,v_1^0}^{\text{pd}}(V_1) = \bg_{h, v_1^0, 1} (V_1)^{\top} \bD_{h, v_1^0, 1}^{-1} \ME[\bg_{h, v_1^0, 1} (V_1) K_{h, v_1^0}(V_1) \theta_1(V_1)].
\]
Hence the un-centered influence function is
\[
\be_1^{\top} \bD_{h, v_1^0, 1}^{-1} [\bg_{h, v_1^0, 1} (V_1) K_{h, v_1^0}(V_1) ({\varphi}_1^{\text{pd}}(\bZ) - \gamma_{h,v_1^0}^{\text{pd}}(V_1)) + \br(\bV_{-1}) ].
\]
One can show that
\[
\be_1^{\top} \bD_{h, v_1^0, 1}^{-1} \ME[\bg_{h, v_1^0, 1} (V_1) K_{h, v_1^0}(V_1) ({\varphi}_1^{\text{pd}}(\bZ) - \gamma_{h,v_1^0}^{\text{pd}}(V_1)) ] = 0
\]
and
\begin{align*}
\ME[\br(\bV_{-1})] & = \int \bg_{h,v_1^0, 1}(v_1) K_{h,v_1^0} (v_1) \tau_v(\bv) f_1(v_1) dv_1 \, f_{-1}(\bv_{-1}) d \bv_{-1} \\
& = \int \bg_{h,v_1^0, 1}(v_1) K_{h,v_1^0} (v_1) \theta_1(v_1) f_1(v_1) d v_1,
\end{align*}
Therefore, the centered influence function of $\theta_h(v_1^0;\MP)$ is
\[
\begin{aligned}
     \phi_{h,v_1^0}^{\text{pd}} (\bZ) = &\, \be_1^{\top} \bD_{h, v_1^0, 1}^{-1} \left[\bg_{h, v_1^0, 1} (V_1) K_{h, v_1^0}(V_1) ({\varphi}_1^{\text{pd}}(\bZ) - \gamma_{h,v_1^0}^{\text{pd}}(V_1)) \right.  \\
    &\, \left. + \br(\bV_{-1}) - \int \bg_{h,v_1^0, 1}(v_1) K_{h,v_1^0} (v_1) \theta_1(v_1) f_1(v_1) d v_1 \right].
\end{aligned}
\]
The efficient influence function of $\be_3^{\top} \bD_{b, v_1^0, 3}^{-1} \ME[\bg_{b, v_1^0, 3} (V_1) K_{b, v_1^0}(V_1) \theta_1(V_1)]$ can be similarly obtained and the result of the theorem follows.

\section{Additional Simulation Results}\label{appendix:addi-simu}

\begin{figure}[htbp]
  \centering
    \begin{subfigure}[t]{0.8\textwidth}
    \includegraphics[width=\textwidth]{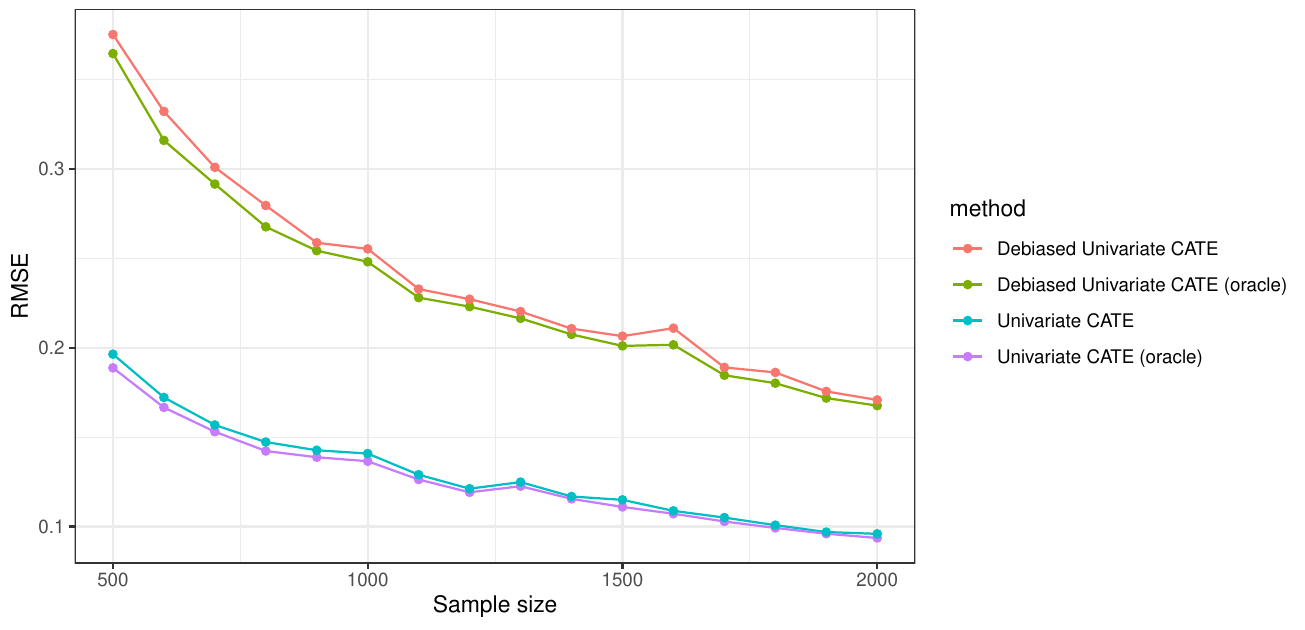}
    \caption{RMSE V.S. Sample size $n$ \label{fig:simu_cate_RMSE_n}}
  \end{subfigure}
    \begin{subfigure}[t]{0.8\textwidth}
    \includegraphics[width=\textwidth]{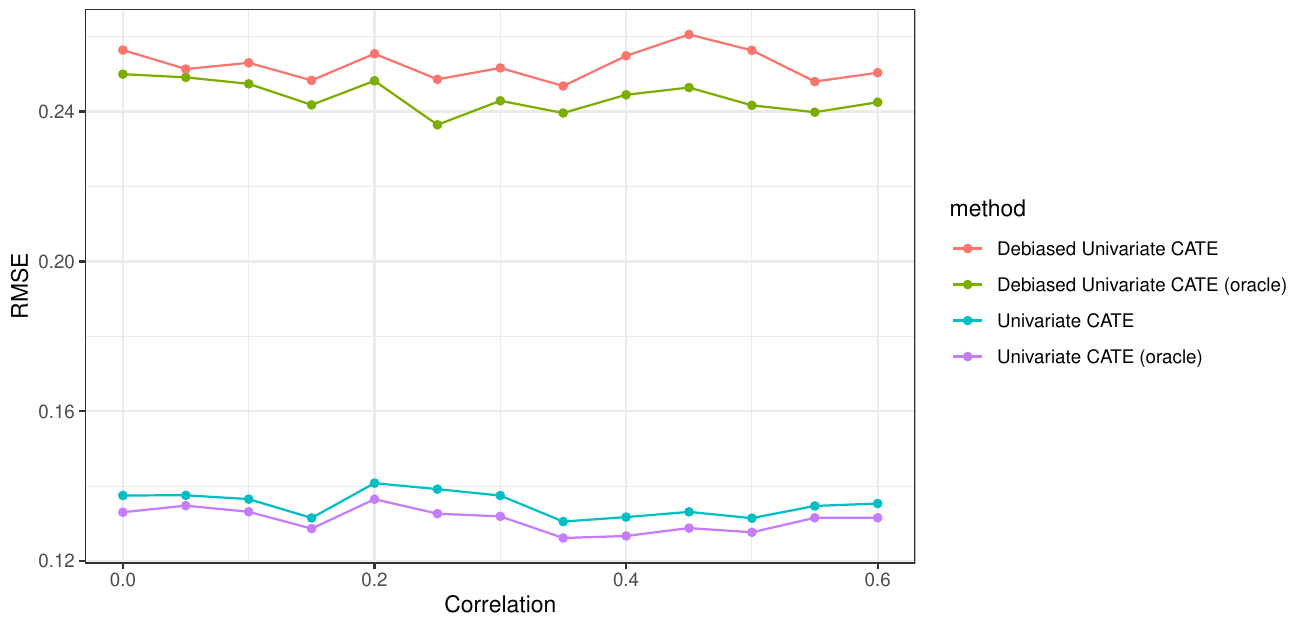}
    \caption{RMSE V.S. Correlation $\rho$ \label{fig:simu_cate_RMSE_rho}}
  \end{subfigure}
  \caption{Simulation results for estimating the univariate CATE function $\tau_1$}
  \label{fig:simu_cate}
\end{figure}

We present simulation results regarding the estimation of univariate CATE curves. This complements our simulation study reported in Section \ref{sec:simulation}. The data generating process and nuisance estimation are described in Section \ref{sec:simulation}. The results are summarized in Figure \ref{fig:simu_cate}. As expected and shown in Figure \ref{fig:simu_cate_RMSE_n}, the estimated RMSE decreases as the sample size increases. One major difference between the univariate CATE estimation and the GAM and partial dependence function estimation is that the estimator of the former is not really sensitive to the correlation between the effect modifiers, as shown in Figure \ref{fig:simu_cate_RMSE_rho}. 

\end{document}